\documentclass[final,3p,times]{elsarticle}

\usepackage{amssymb}
\usepackage{amsmath}
\usepackage{booktabs}
\usepackage{algpseudocode}
\usepackage{longtable}
\usepackage{tabularx}
\usepackage{siunitx}
\usepackage{bbm}
\usepackage{graphicx}
\usepackage{subcaption}
\usepackage{caption}
\usepackage[colorinlistoftodos]{todonotes}
\usepackage{booktabs}
\usepackage{enumitem}
\usepackage{amsfonts}
\usepackage{amssymb}
\usepackage{amsmath}
\usepackage{cancel}
\usepackage{mathtools}
\usepackage{bm}
\usepackage{esvect}
\usepackage{gensymb}
\usepackage{float}
\usepackage{tabulary}
\usepackage{nomencl}
\usepackage{multicol}
\usepackage[hyphens]{url}
\usepackage[hidelinks]{hyperref}
\usepackage[capitalize]{cleveref}

\crefname{appendix}{}{}
\Crefname{appendix}{}{}
\crefformat{appendix}{#2#1#3}
\Crefformat{appendix}{#2#1#3}

\journal{Applied Soft Computing}

\begin{document}

\begin{frontmatter}

\title{Multiobjective optimization-based design and dispatch of islanded hybrid microgrids for remote off-grid communities in sub-Saharan Africa} 

\author[1]{Vineet Jagadeesan Nair} 

\affiliation[1]{
            organization={Department of Mechanical Engineering, Massachusetts Institute of Technology},
            addressline={77 Massachusetts Avenue}, 
            city={Cambridge},
            postcode={02141}, 
            state={MA},
            country={USA}
}

\begin{abstract}

Sizing an islanded hybrid microgrid requires evaluating candidate designs through a year-long hourly simulation. The resulting objective is non-smooth, which limits gradient-based methods and requires approximation for mixed-integer linear programming. We size a solar photovoltaic (PV), wind, lithium-ion battery and diesel microgrid for an off-grid community in Timbila, Kenya, using five objectives: levelized cost of electricity (LCOE), emissions, unmet load, dumped energy and fossil share. We benchmark eight solvers, including particle swarm optimization (PSO), over 30 runs each with a budget of 2,000 evaluations. PSO, the PSO--pattern-search hybrid, pattern search and surrogate optimization finish within 0.5\% of the exhaustively computed integer optimum in all 30 runs; PSO is fastest (expected running time 2.9~s), and gradient-based multistart methods perform worst. We then approximate the Pareto front, select designs with knee-point and compromise rules and the technique for order of preference by similarity to ideal solution (TOPSIS), trace a cost--reliability front over the generator rating, and re-rank candidates over 500 Monte Carlo resource and load scenarios. The equal-weight design, 73~kW of PV with a 239~kWh battery and no wind turbines, supplies 97\% of energy from renewables at 0.184~USD/kWh, 63\% below diesel-only supply. The selected design that serves all load uses a 12.2~kW generator, and accepting 1\% unmet load lowers LCOE to 0.163~USD/kWh. Under uncertainty, a design with a 31\% larger PV array lowers the 95\% conditional value-at-risk of LCOE by 10\% for a 2\% higher expected cost. LCOE is most sensitive to the interest rate over the ranges tested.
\end{abstract}

\begin{highlights}
\item Eight solvers compared over 30 runs each at an equal evaluation budget
\item Pareto designs selected by knee point, compromise and TOPSIS rules
\item Monte Carlo analysis compares expected cost with the risk of high costs
\item PV--battery--diesel design supplies 97\% renewable energy at 63\% lower cost than diesel
\item Reliability targets mainly change generator size; PV capacity changes little
\end{highlights}


\end{frontmatter}

\section*{Nomenclature}

\begin{tabular}{ll}
\multicolumn{2}{l}{\textit{Acronyms}} \\
BS  & Battery storage system \\
CRF & Capital recovery factor \\
DE  & Diesel engine \\
DG  & Distributed generator (backup) \\
DPSP & Deficiency of power supply probability \\
DR  & Demand response \\
DSM & Demand-side management \\
EED & Economic and environmental dispatch \\
GA  & Genetic algorithm \\
LA  & Lead-acid battery \\
LCOE & Levelized cost of electricity (USD/kWh) \\
LI  & Lithium-ion battery \\
MG  & Microgrid \\
MILP & Mixed-integer linear program \\
MPC & Model predictive control \\
MT  & Microturbine \\
PSO & Particle swarm optimization \\
PV  & Solar photovoltaic array \\
REF & Renewable energy fraction (penetration) \\
REPG & Relative excess power generated (dump fraction) \\
RES & Renewable energy sources \\
SA  & Simulated annealing \\
SOC & State of charge \\
SSA & Sub-Saharan Africa \\
TAC & Total annualized cost (USD/y) \\
WT  & Wind turbine \\[6pt]
\multicolumn{2}{l}{\textit{Key variables}} \\
$E_{b,\text{init}}$ & Initial rated battery energy capacity (kWh) \\
$n_s$ & Number of solar PV modules \\
$n_w$ & Number of wind turbines \\
$P_{BS}(t)$ & Battery power output at time $t$ (kW) \\
$P_{DG}(t)$ & Distributed generator power output at time $t$ (kW) \\
$P_{RES}(t)$ & Total renewable power output at time $t$ (kW) \\
$w_i$ & Weight on objective $i$ \\
\end{tabular}

\section{Introduction and problem background \label{sec:access}}

Electrification rates in sub-Saharan Africa (SSA) remain low \cite{SSA_decent}. In 2023, 666 million people, about 8\% of the world's population, still lacked access to electricity, and 565 million of them lived in SSA \cite{sdg7_2025}. Up to 85\% of the unserved population lives in remote rural villages far from national grids, and many of the remaining 15\%, who live closer to a grid, cannot afford the connection, metering and wiring costs \cite{quak_CBA}. Electricity access is closely tied to living standards, productivity, employment and health, so improving it supports poverty alleviation and several of the United Nations Sustainable Development Goals (SDGs) \cite{SSA_decent}.

Connecting isolated communities to the grid is expensive. According to Power for All, grid extension and infrastructure upgrades in sub-Saharan Africa can cost up to \$5.5 million per MW/km \cite{grid_gtm}, and long planning, procurement, environmental assessment and construction phases lead to budget overruns and delays. Connection fees average about \$136 for a single-phase connection and can exceed monthly household incomes, so extension rarely pays in sparsely populated rural areas with low, dispersed demand. Even households with access often spend a large share of their income on electricity and fuel. Close to half of the world's population still relies on firewood or charcoal for cooking and heating and on kerosene for lighting, which are expensive and harm health, and almost two-thirds of the world's poorest people live in rural areas \cite{kammen}. Rural electrification therefore needs affordable, reliable options that can be deployed quickly.

\subsection{Microgrids and their role in rural electrification}

Microgrids are low-voltage, local systems of generation, storage and demand. A grid-connected microgrid exchanges power with the main grid, within transmission limits, and can disconnect at the point of common coupling to operate on its own; its operator can also sell surplus renewable power to the grid. An islanded microgrid has no such connection: it must meet demand locally and manage surplus generation through storage, curtailment or dump loads. The growth of distributed generation from renewable sources (RES) such as wind, solar, geothermal, biomass and small hydro has made both types more practical. Because grid extension is costly (\cref{sec:access}), this work considers only stand-alone systems that always operate in islanded mode, which adds the challenges discussed in \cref{sec:MG_challenges}.

\subsection{Advantages of microgrids}

Microgrids are much smaller than thermal and nuclear baseload plants but more modular and flexible, which eases the integration of variable distributed energy resources (DERs) that are often located far from load centers. Feeding variable renewables into a conventional grid can degrade power quality through unsteady voltages and harmonics, and large interconnected grids are vulnerable to natural disasters, extreme weather and cyber-physical attacks: a fault in one section can cascade, as the blackouts, brownouts and load-shedding common in urban areas of developing nations show \cite{macrogrid_prob}.

By displacing fossil fuels, microgrids can lower greenhouse gas emissions and support the SDGs and the Paris Agreement. Because they can island during faults and emergencies, they keep supplying local load and make the wider network more resilient, and controllable microgrids can relieve congestion and provide ancillary services to the main grid, such as load shifting, frequency or voltage control and peak shaving. Generating close to demand also reduces transmission and distribution (T\&D) losses and defers T\&D investment. Microgrids are among the few realistic options for low-cost local supply at other remote, resource-constrained sites, such as islands, military bases and Arctic settlements. The International Energy Agency (IEA) estimates that about 70\% of the rural areas that lack access will be connected through mini-grid, microgrid or off-grid solutions \cite{WEO_2018,mentis}, which could bring electricity to about 140 million rural Africans through 100,000 mini-grids by 2040 \cite{wri}.

\subsection{Challenges with microgrids \label{sec:MG_challenges}}

Small microgrids lack the averaging effect of large networks that connect regions with different weather and load patterns. Their relative load fluctuations are therefore larger, and unfavorable weather, such as extended wind lulls or cloudy periods, can leave demand unmet. To cover worst cases, designers often oversize renewable and storage capacity, which raises project cost.

Isolated microgrids have no main grid, with its peaker plants and interconnectors, to fall back on, and they lack large synchronous generators whose inertia resists sudden changes in frequency and voltage. Mismatches between supply and demand of real and reactive power must therefore be corrected quickly to keep frequency and voltage stable \cite{tech_challenges}. This study addresses power balancing on time scales of hours to months. Frequency and voltage stability act over seconds to minutes and depend mainly on switchgear and power-electronic design, so they are outside its scope.

\section{Literature review}

Research on microgrids ranges from the selection and sizing of generation and storage technologies to the real-time energy management and dispatch of these components.

\subsection{Sizing optimization}

Once the technologies have been chosen on economic, technical and socio-political grounds, sizing sets the rating and number of units of each renewable source (for example, the number of wind turbines (WT) or the area of the solar photovoltaic (PV) array), the power rating and energy capacity of storage, and the rated power of backup diesel engines or gas turbines. Most studies of stand-alone systems consider \textit{hybrid} designs that combine two or more renewable sources with fossil-fueled distributed generators (DGs) as backup to meet reliability requirements \cite{kaabeche2014} \cite{bhandari2014}. They minimize economic cost \cite{tazvinga2013}, pollutant emissions and measures of power quality, reliability and efficiency \cite{kaabeche2011a}, which reflect the goals of the community served. \cref{app:costs} details the individual cost terms.
 
Past studies show the trade-offs involved. PV/WT systems that use both diesel generators and batteries are more economical than those that use only batteries or only diesel engines, and in a region of Algeria with a strong solar resource, configurations with far fewer wind turbines than PV modules had the lowest energy cost and net CO$_2$ emissions \cite{kaabeche2014}; Timbila, Kenya, the site studied here, also has a strong solar resource. Sizing results therefore depend on the local climate, the wind and solar resources and the demand profile, and also on the objectives and on financial parameters such as interest, discount and inflation rates, because capital dominates the cost of renewables and storage while fuel dominates that of backup generation. The objectives often compete, since more reliable and cleaner systems usually cost more; matching the reliability of the main grid can make off-grid PV--battery systems up to ten times more expensive than on-grid solutions \cite{weng2016}.

Models of hybrid microgrids combine well-established models of the wind, solar, diesel, gas and battery subsystems \cite{bhandari2014}. The sizing of storage and backup generation often determines system performance: the efficiency of microgrid operation depends strongly on battery scheduling, because the battery shifts peak load and smooths fluctuations in demand and renewable supply \cite{gu2014}. Most studies that build their own models simulate one year (8,760~h) \cite{logenthiran2010}, while commercial and open-source design tools can simulate longer periods; \cref{app:tools} reviews the main tools.

\subsection{Optimal dispatch}

Operational studies address unit commitment, which decides which dispatchable units (generators and storage) are online in each period to meet forecast demand at minimum cost \cite{saravanan2013}, and unit dispatch, which sets the output of each online unit in each interval.

Single-objective formulations usually minimize cost, but multiobjective formulations that also cover emissions, power loss and heat are increasingly common \cite{pathak2017}. Combined economic and emissions dispatch problems are generally nonlinear, non-smooth and non-convex, with demanding equality and inequality constraints, which makes conventional optimization methods ill-suited and raises the risk of convergence to local minima \cite{pathak2017}. As mixed-integer problems they combine discrete and continuous decision variables in high dimension, which rules out disciplined convex programming tools such as CVX and has motivated metaheuristic approaches. Some studies replace hard constraints by heavily penalized soft constraints to keep the problem feasible \cite{zachar2016}.

Other operational research covers seamless islanding and reconnection to the grid \cite{mg_review}, primary and secondary control of frequency and voltage, and tertiary control for economic optimization \cite{mg_review}. Microgrid modeling can also draw on models of electric vehicles as controllable loads, weather-dependent generation such as solar PV, controllable generation and storage \cite{mg_mod_opt}. Some studies simulate simplified dynamic models of small microgrids in MATLAB \texttt{Simulink}.

Temporal patterns in load and renewable supply, from hourly to seasonal scales, determine how the battery cycles and how long the backup DG runs, and hence fuel use and operating cost \cite{tazvinga2013}. Optimal configuration and operation improve power quality, reliability, security and sustainability and lower the cost to consumers \cite{pathak2017}. For islanded microgrids without grid support, incentive-based demand response (DR) \cite{nwulu2017} and model predictive control \cite{zachar2016} help reject forecast errors and noise in load and renewable output. Game-theoretic work on demand-side management (DSM) shows that price signals that lead individual agents to minimize their own costs can produce globally optimal aggregate demand profiles \cite{paola2017}. Integrated design and operation optimization, which sizes and dispatches the system in a single stage rather than in two sequential steps, is another recent direction \cite{moshi2016}.

\subsection{Survey of optimization algorithms}

Many studies formulate sizing and dispatch as nonlinear mixed-integer programs solved with global heuristic methods; others use MILP relaxations for computational tractability \cite{parisio2014, Qu2018}. The seven algorithms benchmarked in this work (\cref{sec:results}) are summarized below; full derivations and parameter details are in \cref{app:opt_alg}. Genetic algorithms (GA) and evolutionary algorithms (EA) are stochastic, population-based methods that search the solution space by mimicking biological evolution through selection, mutation, and crossover of candidate solutions \cite{nemati2015}. They require no gradient information, handle integer and mixed-integer variables natively, and can generate full Pareto-optimal solution sets in a single run \cite{fadaee2012}. Their main drawback is that performance depends heavily on tuning of population size, crossover rate, and mutation rate, and complexity grows with the number of decision variables \cite{fathima2015}. Particle swarm optimization (PSO) simulates the social behavior of a swarm: each particle moves through the search space guided by its own best-known position and the swarm's global best, updating velocity at each iteration \cite{fathima2015, pathak2017}. It is generally easier to implement than GA, has fewer hyperparameters, converges quickly on non-convex problems, and has been applied successfully to microgrid sizing and economic dispatch \cite{alrashidi2010}. It can struggle with equality constraints and may stagnate in flat regions of the objective landscape. The velocity-update equations are given in \cref{app:pso}.

Surrogate optimization (SO) constructs a radial-basis-function interpolant of the objective using quasirandom sample points, then minimizes the surrogate rather than the expensive true function \cite{gutmann2001}. This is advantageous for computationally costly evaluations, and the method is proven to converge to the global optimum on bounded domains. In practice, however, refitting the surrogate adds a large overhead per evaluation, which dominates the run time when the true objective is cheap, as here (\cref{tab:bench_stats}), and it lacks an explicit stopping criterion based on optimality. Simulated annealing (SA) is a single-solution method that accepts worse solutions with a temperature-dependent probability, allowing it to escape local optima in analogy with the physical annealing of metals \cite{fathima2015}. It is theoretically guaranteed to converge globally under a logarithmic cooling schedule, but this schedule is extremely slow in practice. SA also does not support parallel computation. MultiStart (MS), GlobalSearch (GS) and direct pattern search (DPS) are gradient-based or direct-search local methods run from multiple starting points to approximate global solutions. MS and GS use local gradient solvers (\texttt{active-set}) launched from many random starts, which makes them faster than population methods on smooth landscapes but less reliable on non-smooth or flat objectives. DPS uses a mesh-based direct search with no gradient information, which is more robust to non-smoothness but can need many evaluations when the mesh must be refined; on our problem it reached the target in every run but used nearly the full evaluation budget (\cref{tab:bench_stats}).

Recent work extends these methods in three directions relevant here. Improved multiobjective variants of evolutionary and swarm algorithms return Pareto sets rather than single designs, for example an improved NSGA-II for the dispatch of hybrid ship power systems \cite{wang2024nsga} and an improved multiobjective grey wolf optimizer for sizing an island microgrid \cite{zhu2020imogwo}; grey wolf optimization has also been compared with PSO for wind--solar sizing \cite{geleta2022gwo}, and hybrids of PSO with other heuristics aim to balance exploration and exploitation \cite{javidrad2017psosa}. Several studies pair multiobjective search with multi-criteria decision making to choose one design from the Pareto set, such as multiobjective differential evolution followed by the analytic hierarchy process (AHP) and the technique for order of preference by similarity to ideal solution (TOPSIS) for stand-alone PV sizing \cite{muhsen2019mode} and weighted-sum Pareto sets ranked with ELECTRE for microgrid scheduling \cite{karimi2019madm}. Resource and load uncertainty is also increasingly propagated into the design stage, through Markov chain Monte Carlo scenario generation for stand-alone microgrid planning \cite{bakhtiari2021mcmc} or two-stage stochastic programming for off-grid hybrid systems \cite{calabrese2026tssp}. For stochastic solvers, the computational-intelligence literature recommends comparing result distributions over repeated independent runs with nonparametric tests rather than single runs \cite{derrac2011tutorial,carrasco2020trends}; we follow that practice in \cref{sec:results}.

\subsection{Control strategies and design tools \label{sec:control_survey}\label{app:design_tools}}
Real-time microgrid control (hierarchical primary, secondary and tertiary levels; centralized, decentralized and distributed architectures; model predictive control) lies outside the scope of this paper and is surveyed in \cref{app:control}. Widely used design tools (HOMER Pro, DER-CAM, REopt and the Microgrid Design Toolkit) are compared in \cref{app:tools}.

\subsection{Research gaps and contributions}

Three gaps motivate this work. First, most multiobjective microgrid studies treat grid-connected systems that island only under faults; permanently isolated systems, which must meet all load from local resources, receive less attention. Second, comparisons of stochastic solvers for microgrid sizing often rest on one run per solver, so differences between solvers cannot be separated from run-to-run variation. Third, studies that compute Pareto sets often stop short of a rule for choosing one design from them, and robustness is commonly checked with a few deterministic scenarios rather than by propagating resource and load uncertainty.

Sub-Saharan African conditions shape the parameter ranges and the design questions in this study. Diesel prices in remote areas are high and volatile (our default of 3.20 USD per US gallon, about 0.85 USD/L, is conservative for remote sites, and \cref{sec:sizing_sensitivity} spans 0.2--12 USD/gal); nominal interest and inflation rates are high (9\% and 5.7\% for Kenya in the base case), which penalizes capital-intensive storage; measured climate data are scarce, so we draw on the TAHMO station network and check it against reanalysis (\cref{app:climate}); and grid extension is expensive, which motivates the break-even distance analysis in \cref{sec:results}.

This paper makes four contributions. (i) We formulate the sizing of an islanded PV--wind--battery--diesel microgrid as a simulation-based, non-smooth optimization problem in which an 8,760-hour rule-based operating simulation scores each candidate on five normalized objectives, and we state how integrality, normalization and reliability enter the formulation. (ii) We benchmark seven global optimization solvers and a PSO--pattern-search hybrid on this problem at an equal budget of 2,000 evaluations over 30 independent runs each, using success rates against the integer optimum found by exhaustive enumeration, expected running time and nonparametric tests. (iii) We pair a priori weighted-sum sizing with a posteriori Pareto search, select designs from the Pareto set with knee-point, compromise and TOPSIS rules, and re-rank the candidates by expected cost, conditional value-at-risk of cost and loss-of-load probability over 500 Monte Carlo resource and load scenarios. (iv) We apply this workflow to an off-grid community in Timbila, Kenya, with fuel price, interest, inflation and storage cost ranges drawn from sub-Saharan African conditions, and identify which design rules carry over to other sites once local inputs are recalibrated. \cref{tab:litcomp} places the work relative to representative prior studies.

\begin{table}[htbp]
\centering
\footnotesize
\setlength{\tabcolsep}{3pt}
\caption{\label{tab:litcomp} Comparison of representative prior studies on multiobjective hybrid microgrid optimization. \textbf{Obj}: objectives (C = cost, E = emissions, R = reliability, D = dump/curtailment; multi = several, not detailed in the abstract). \textbf{Isl.}: islanded/off-grid. \textbf{Size}/\textbf{Disp.}: sizing/dispatch performed. \textbf{R/S}: robustness or sensitivity analysis. \textbf{Sel.}: explicit rule for selecting a design from a Pareto set. \textbf{UQ}: probabilistic uncertainty propagation. \textbf{Stats}: repeated-run statistical comparison of solvers. \textbf{SSA}: sub-Saharan Africa case.}
\begin{tabular}{>{\raggedright\arraybackslash}p{2.8cm} >{\raggedright\arraybackslash}p{1.1cm} >{\raggedright\arraybackslash}p{2.9cm} cccccccc}
\toprule
\textbf{Study} & \textbf{Obj.} & \textbf{Method} & \textbf{Isl.} & \textbf{Size} & \textbf{Disp.} & \textbf{R/S} & \textbf{Sel.} & \textbf{UQ} & \textbf{Stats} & \textbf{SSA} \\
\midrule
Kaabeche et al.~\cite{kaabeche2014} & C, E & Iterative & \checkmark & \checkmark & & & & & & \\
Bhandari et al.~\cite{bhandari2014} & C & HOMER & & \checkmark & & & & & & \\
Tazvinga et al.~\cite{tazvinga2013} & C, E & GA & \checkmark & \checkmark & \checkmark & & & & & \checkmark \\
Logenthiran et al.~\cite{logenthiran2010} & C & GA & \checkmark & \checkmark & \checkmark & & & & & \\
Moshi et al.~\cite{moshi2016} & C & GA & \checkmark & \checkmark & \checkmark & & & & & \checkmark \\
Zachar \& Daoutidis~\cite{zachar2016} & C & MPC & & & \checkmark & \checkmark & & & & \\
Parisio et al.~\cite{parisio2014} & C & MILP & \checkmark & & \checkmark & & & & & \\
Nwulu \& Xia~\cite{nwulu2017} & C, E & DR/MPC & & & \checkmark & & & & & \\
Weng et al.~\cite{weng2016} & C & DER-CAM & & \checkmark & & & & & & \\
Gu et al.~\cite{gu2014} & C & MPC & \checkmark & & \checkmark & & & & & \\
Karimi \& Jadid~\cite{karimi2019madm} & multi & Weighted sum + ELECTRE & & & \checkmark & & \checkmark & & & \\
Muhsen et al.~\cite{muhsen2019mode} & C, R & MODE + AHP--TOPSIS & \checkmark & \checkmark & & & \checkmark & & & \\
Zhu et al.~\cite{zhu2020imogwo} & C, R & Improved MOGWO & \checkmark & \checkmark & & & & & & \\
Bakhtiari et al.~\cite{bakhtiari2021mcmc} & multi & MCMC scenarios & \checkmark & \checkmark & & & & \checkmark & & \\
Wang et al.~\cite{wang2024nsga} & multi & Improved NSGA-II & \checkmark & & \checkmark & & & & & \\
Calabrese et al.~\cite{calabrese2026tssp} & C, E & Two-stage SP, $\varepsilon$-constraint & \checkmark & \checkmark & \checkmark & \checkmark & & \checkmark & & \\
\textbf{This work} & C, E, R, D & Seven solvers + PSO--PS; \texttt{paretosearch}; TOPSIS & \checkmark & \checkmark & \checkmark & \checkmark & \checkmark & \checkmark & \checkmark & \checkmark \\
\bottomrule
\end{tabular}
\end{table}

\section{Microgrid system description \label{chap:methods}}

\begin{figure}[htbp]
  \centering
  \includegraphics[width=0.7\linewidth]{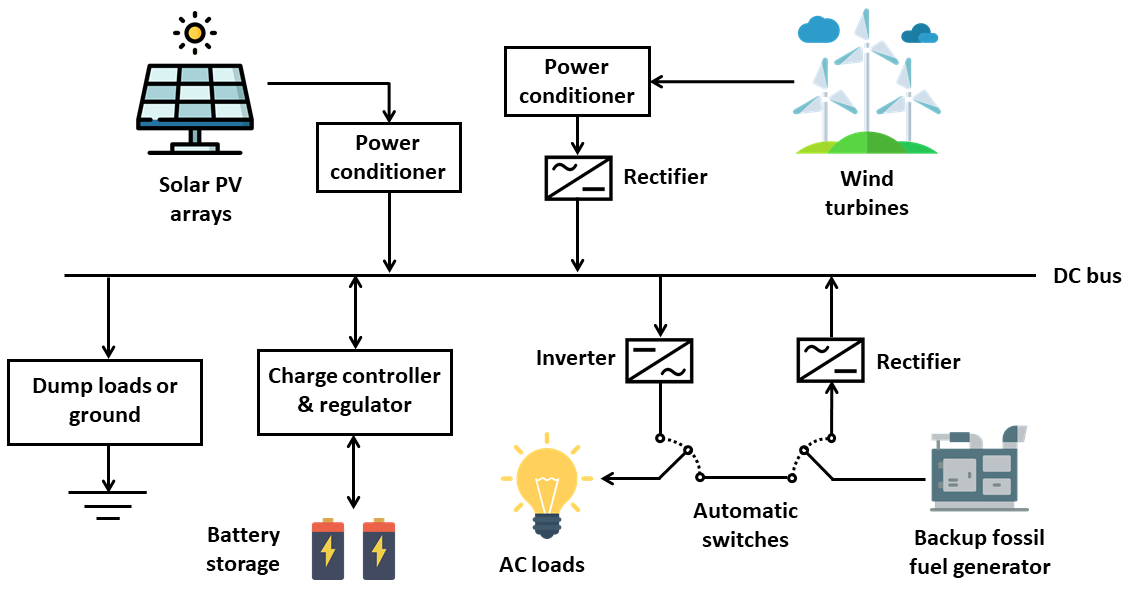}
  \caption{Schematic illustrating key components and power flows in the hybrid microgrid.}
  \label{fig:chap3_MG}
\end{figure}

The microgrid (\cref{fig:chap3_MG}) consists of:

\begin{itemize}
    \item non-dispatchable renewable sources (RES): solar photovoltaic arrays (PV) and wind turbines (WT);
    \item dispatchable fossil-fueled distributed generators (DG) for backup: a diesel engine (DE) or a microturbine (MT);
    \item battery storage (BS): lithium-ion (LI) or lead-acid (LA);
    \item charge controllers that regulate the battery's state of charge (SOC), charging and discharging;
    \item power conditioners that regulate the power delivered from the RES to the DC bus;
    \item bidirectional converters that act as an inverter (DC to AC) or a rectifier (AC to DC).
\end{itemize}

The system is a hybrid, islanded microgrid that serves a remote rural community: it can neither buy power from nor sell power to an external grid. \cref{app:mg_models} gives the component models, the climate data for the site in Kenya and the load characteristics.

\section{Optimization problem formulation \label{sec:opt_prob}}

\begin{figure}[htbp]
  \centering
  \includegraphics[width=\linewidth]{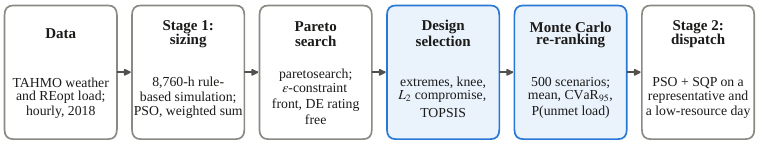}
  \caption{\label{fig:workflow} Design workflow. Stage~1 sizes the system under a rule-based controller; Pareto search, rule-based selection from the front and Monte Carlo re-ranking choose among candidate designs; Stage~2 optimizes day-ahead dispatch of a selected design on a representative and a low-resource day and tests it on stress days.}
\end{figure}

The optimization proceeds in two stages (\cref{fig:workflow}). In Stage~1 (sizing), an annual hourly simulation with a rule-based dispatch inner loop is embedded in an outer global optimization loop that chooses the PV, WT and battery capacities. In Stage~2 (dispatch), the sized system is taken as given and a multiobjective economic-environmental dispatch problem is solved for a representative day and for a low-resource day. The Stage-1 controller is non-anticipative: it uses only the current hour's load, renewable output and state of charge, and it follows the load-following logic of most deployed off-grid controllers and of HOMER Pro's default strategy. Each candidate design is therefore scored under an operating policy that an operator can implement without forecasts. Embedding a 24-hour dispatch optimization inside every sizing evaluation would multiply the cost of each of the thousands of evaluations per run by 365 daily solves. A design ranked best under rule-based operation may rank differently under optimized dispatch. \cref{sec:limitations} discusses this limitation, and \cref{sec:dispatch_opt} reports how optimized dispatch changes daily cost and emissions for the selected design.

The two stages are:

\begin{enumerate}
    \item \textbf{Design and sizing (Stage 1)} over one year (8,760~h) in hourly steps ($\Delta t = 1$~h). The decision variables are the number of PV modules $n_s$, the number of wind turbines $n_w$ and the initial rated battery capacity $E_{b,\;init}$. Inside the optimizer, the annual simulation dispatches the units by a priority rule (\cref{sec:disp_strat}) rather than by an inner optimization, which keeps each evaluation fast: one evaluation of the 8,760-hour simulation takes about 2\,ms on a current laptop, so a sizing run of 2,000 evaluations completes in seconds (\cref{sec:results}).
    \item \textbf{Day-ahead dispatch (Stage 2)}, an economic and environmental dispatch (EED) over 24~h in hourly steps. With the capacities fixed by Stage 1, it optimizes the DG output $P_{DG}(t)$ and the battery power $P_{BS}(t)$; the renewable output $P_{RES}$ follows the weather and is not dispatchable.
\end{enumerate}

The problem has five objectives: (i) levelized cost of electricity (LCOE), (ii) emissions (Em), (iii) deficiency of power supply probability (DPSP), (iv) relative excess or dump power generated (REPG) and (v) renewable energy fraction (REF). This section states the optimization problems and their constraints; \cref{app:costs} defines the costs and objectives.

\subsection{Dispatch strategy \label{sec:disp_strat}}
In the sizing problem, a priority order for the generation and storage units is encoded in the objective function. \cref{tab:dispatch_strat_for_sizing} lists this load-following rule.
\begin{table}[htbp]
\centering
\begin{tabular}{p{0.75cm} p{13cm}}
\toprule
\multicolumn{1}{c}{\textbf{Step}} & \multicolumn{1}{c}{\textbf{Action}}    \\
\midrule
1 & Use the RES output to meet the load. \\ [1ex]
2 & If the RES output (PV and WT) exceeds the load, charge the battery with the surplus. \\ [1ex]
3 & If the surplus exceeds what the battery can absorb (its power rating or the headroom below $SOC_{max}$), send the remainder to the dump load. \\ [1ex]
4 & If the RES output falls short of the load, discharge the battery. \\ [1ex]
5 & If the battery reaches $SOC_{min}$ or its power rating and load remains unmet, dispatch the backup DG until all demand is met or the DG reaches its rated power. \\ [1ex]
6 & The DG cannot run below its minimum stable output $P_{DG,\;min}$. If the remaining deficit in step 5 is smaller, the DG runs at $P_{DG,\;min}$ and the surplus is rectified from AC to DC and stored in the battery or dumped, depending on its SOC (step 3). \\ [1ex]
7 & Any demand still unmet is recorded as lost load and counts toward DPSP. Incentive-based demand response (DR) or demand-side management (DSM), which ask consumers to shift or curtail load, are an alternative to load shedding (\cref{sec:dr}). \\ [1ex]
\bottomrule
\end{tabular}
\caption{Load-following rule used in the sizing simulation.}
\label{tab:dispatch_strat_for_sizing} 
\end{table}

\subsection{Constraints}

\begin{table}[htbp] 
\centering
\begin{tabular}{p{5cm} p{10cm}}
\toprule
\multicolumn{1}{c}{\textbf{Constraint}} & \multicolumn{1}{c}{\textbf{Description}}    \\
\midrule
Power limits of the DG & When the DG is online ($P_{DG}(t) > 0$), its output lies between its minimum stable output and its rated power. \\ [1ex]
Positivity & The DG output is non-negative. \\ [1ex]
Ramp rate limits on the DG & Not enforced: NERC (North American Electric Reliability Corporation) disturbance control standards require regulating and supplemental reserves to reach full nameplate capacity within 10--20 minutes \cite{nerc}, well within the hourly time step. \\ [1ex]
Power limits on the BS & Charging (negative) and discharging (positive) battery power do not exceed the power rating. \\ [1ex]
Safe battery operation & The SOC stays between $SOC_{min}$ and $SOC_{max}$. \\ [1ex]
Reliability requirement & DPSP does not exceed an upper limit (dispatch problem). \\ [1ex]
\bottomrule
\end{tabular}
\caption{Constraints of the sizing and dispatch problems.}
\label{tab:dispatch_constraints} 
\end{table}

\cref{tab:dispatch_constraints} lists the constraints. The weighted sizing problem is:
\begin{equation}
    \label{eq:sizing}
    \min_{n_s, \; n_w, \; E_{b,\;init}} w_1 \cdot \frac{LCOE}{LCOE_{base}} \; + \; w_2 \cdot \frac{Em}{Em_{base}} \; + \; w_3 \cdot DPSP \; + w_4 \; \cdot REPG \; + \; w_5 \cdot (1-REF)
\end{equation}
where the weights sum to one ($w_1 + w_2 + w_3 + w_4 + w_5 = 1$) and $n_s, \; n_w, \; E_{b,\; init} \geq 0$. The remaining constraints are enforced inside the MATLAB simulation that evaluates the objective, through the dispatch rule of \cref{sec:disp_strat}.

Eq.~(\ref{eq:sizing}) is a weighted-sum scalarization of the five objectives. For strictly positive weights its minimizer is Pareto-optimal, but the method can only reach supported Pareto points, those on the convex hull of the Pareto front. We therefore use it as an a priori method for a decision-maker with stated preferences (equal weights by default; \cref{app:weights} varies them) and complement it with the a posteriori Pareto analysis of \cref{sec:pareto_opt,sec:pareto_select}. Reliability enters through the DPSP objective. With the default 16~kW DG the rule-based controller meets all load, so DPSP is zero for every candidate design; \cref{sec:pareto_rel} therefore also treats the DG rating as a decision variable to expose the cost--reliability trade-off.

The day-ahead dispatch problem is:
\begin{equation}
    \label{eq:dispatch}
    \min_{P_{DG}, \; P_{BS}} w_1 \cdot \frac{COE}{COE_{base}} \; + \; w_2 \cdot \frac{Em}{Em_{base}} \; + w_3 \; \cdot REPG \; + \; w_4 \cdot (1-REF)
\end{equation}
subject to the following constraints:
\begin{gather*}
\vv{0}_{1\times24} \; \leq \; \vv{\bm{P_{DG}}}(t) \; \leq \; P_{DG,\;r}\cdot \vv{\mathbbm{1}}_{1\times24}, \quad P_{DG,\;min} \cdot \mathbbm{1}\{P_{DG}(t) > 0\} \; \leq \; P_{DG}(t) \; \leq \; P_{DG,\;r} \\
SOC_{min}\cdot \vv{\mathbbm{1}}_{1\times25} \; \leq \;\vv{\bm{SOC}}(t) \;\leq\; SOC_{max}\cdot \vv{\mathbbm{1}}_{1\times25}, \quad -P_{BS,\;r}\cdot \vv{\mathbbm{1}}_{1\times24} \;\leq \;\vv{\bm{P_{BS}}}(t) \; \leq \; P_{BS,\;r}\cdot \vv{\mathbbm{1}}_{1\times24} \\
DPSP \; \leq \;DPSP_{max}, \quad \Sigma_{i=1}^{i=4} \; w_i  = 1, \; \mathbbm{1}\{P_{DG}(t) > 0\} = 
\begin{cases} 
    1 & \text{if } P_{DG}(t) > 0 \; \text{i.e., } DG_{ON}(t) = 1\\
    0 & \text{otherwise}
\end{cases} 
\end{gather*}

where $\vv{\bm{P_{DG}}}$ and $\vv{\bm{P_{BS}}}$ are $1\times24$ vectors of hourly DG and battery power, and $\vv{\bm{SOC}}$ is a $1\times25$ vector whose last element, $SOC(25)$, keeps the state of charge within limits at the end of the day. $DPSP_{max}$ sets the required reliability as the largest allowed share of daily unmet load. Because each objective is normalized (\cref{app:costs}), the objective values are dimensionless: DPSP, dump and $1-REF$ lie between 0 and 1, while normalized LCOE and emissions exceed 1 only for a design that costs more or emits more than the diesel-only baseline. By default both problems weight the objectives equally: $w_i = 0.2$ in sizing and $w_i = 0.25$ in dispatch.

\subsection{Mathematical properties of the model}

The bounds on power and SOC are linear, but the conditional minimum-output constraint on the DG is nonconvex. The objectives are also nonconvex, nonlinear and non-smooth: discrete states such as the DG status $DG_{ON}$ and the battery cycle count $BS_{cycled}$ make them discontinuous and non-differentiable, and several component sub-models are nonlinear. The numbers of PV modules $n_s$ and wind turbines $n_w$ are integers. Our sizing runs search an integer-coded space: the solvers move within continuous bounds, and the objective rounds $n_s$ and $n_w$ to the nearest integers before simulating, so every evaluated design can be built, and a final one-dimensional search re-optimizes $E_{b,\;init}$ for the rounded design. \cref{sec:results} also reports the alternative relax-and-round strategy, in which the solvers optimize the continuous relaxation and only the returned point is rounded. Mixed-integer formulations of such problems are NP-hard \cite{parisio2013}. The number of variables and constraints of the dispatch problem grows with the horizon and with finer time steps, so we restrict dispatch to one day ahead in hourly steps. Both problems are solved with MATLAB's Optimization and Global Optimization Toolboxes.

\section{Results and discussion \label{sec:results}}

All results were obtained with MATLAB R2026a on an Apple M1 Max with 64~GB of memory, without parallel evaluation.

\subsection{Comparison of solution algorithms for MG planning and design optimization}

The sizing objective is non-smooth: the hourly simulation switches the DE on and off, counts battery cycles and schedules replacements, so the objective has steps and kinks even when $n_s$ and $n_w$ are relaxed to continuous values, and gradient-based solvers can stall on flat or stepped regions. We therefore compare global solvers from MATLAB's Global Optimization Toolbox on the base LI+DE problem (16~kW DE, equal weights $w_i = 0.2$). The comparison follows standard practice for benchmarking stochastic optimizers \cite{hansen2021coco, derrac2011tutorial}: every solver receives the same budget of 2,000 objective evaluations, counted by a wrapper around the objective, and runs 30 times with independent random seeds (\cref{tab:protocol}). Solvers use MATLAB's default settings except for the changes listed in \cref{tab:protocol}. The integer variables are handled by integer coding: $n_s$ and $n_w$ are rounded inside the objective, so every evaluated point is a feasible design. After the search, a local polish evaluates the four integer neighbors of the best point and re-optimizes $E_{b,\;init}$ for each with Brent's method; these evaluations (at most 80) count toward the budget. The bounds are $[0, 0, 0]$ and $[400, 40, 400]$.

\begin{table}[htbp]
\centering
\scriptsize
\begin{tabular}{p{2.6cm} p{2.3cm} p{7.4cm}}
\toprule
\textbf{Solver} & \textbf{MATLAB function} & \textbf{Settings (defaults unless stated)} \\
\midrule
Particle swarm (PSO) & \texttt{particleswarm} & swarm size 30; stops at the budget or on its default stall criterion \\
Genetic algorithm (GA) & \texttt{ga} & population 50; stops at the budget or on default criteria \\
Simulated annealing (SA) & \texttt{simulannealbnd} & random start point; \texttt{MaxFunctionEvaluations} = budget \\
Surrogate optimization (SO) & \texttt{surrogateopt} & \texttt{MaxFunctionEvaluations} = budget \\
MultiStart (MS) & \texttt{MultiStart} + \texttt{fmincon} & active-set local solver; up to 1,000 random start points; stopped at the budget \\
GlobalSearch (GS) & \texttt{GlobalSearch} + \texttt{fmincon} & active-set local solver; 400 trial points, 100 stage-one points \\
Pattern search (DPS) & \texttt{patternsearch} & complete poll; restarted from random points until the budget is spent \\
PSO + pattern search & \texttt{particleswarm}, \texttt{patternsearch} & PSO for 80\% of the budget, then pattern search from the best point \\
\midrule
All & & budget 2,000 evaluations including the integer polish; 30 seeds (1--30); integer-coded $n_s, n_w$; serial evaluation; R2026a on an Apple M1 Max (64~GB) \\
\bottomrule
\end{tabular}
\caption{\label{tab:protocol} Benchmark protocol.}
\end{table}

We score each run by the objective of its final integer design and compare it with the best integer design $f^* = 0.10892$, found by exhaustive enumeration of all $401 \times 41$ combinations of $n_s$ and $n_w$ with $E_{b,\;init}$ optimized for each (grid search followed by Brent's method). A run succeeds if it ends within 0.5\% of $f^*$. The expected running time (ERT) is the total number of evaluations over all runs divided by the number of successful runs \cite{hansen2021coco}; it rewards solvers that reach the target consistently with few evaluations, and is infinite for a solver that never succeeds. We test for differences between solvers with the Kruskal--Wallis test and then compare each solver with the best-ranked one using two-sided Mann--Whitney tests with Holm correction and the Vargha--Delaney effect size $\hat{A}_{12}$ \cite{derrac2011tutorial, vargha2000a12}.

\begin{table}[htbp]
\centering
\scriptsize
\setlength{\tabcolsep}{4pt}
\begin{tabular}{lrrrrrrrr}
\toprule
 & \multicolumn{5}{c}{Gap to $f^*$ [\%]} & & \multicolumn{2}{c}{ERT} \\
\cmidrule(lr){2-6}\cmidrule(lr){8-9}
Solver & Best & Median & Mean & SD & Worst & Success & [evals] & [s]\\
\midrule
PSO+PS & 0.000 & 0.001 & 0.002 & 0.004 & 0.016 & 100\% & 1,575 & 3.04\\
PSO & 0.000 & 0.001 & 0.002 & 0.004 & 0.016 & 100\% & 1,518 & 2.93\\
DPS & 0.000 & 0.001 & 0.004 & 0.004 & 0.017 & 100\% & 1,984 & 3.87\\
SO & 0.000 & 0.007 & 0.009 & 0.008 & 0.032 & 100\% & 1,972 & 24.01\\
GA & 0.027 & 2.890 & 3.813 & 4.904 & 23.002 & 23\% & 8,551 & 16.55\\
SA & 0.007 & 3.264 & 13.536 & 17.873 & 56.888 & 30\% & 6,384 & 12.62\\
GS & 0.078 & 11.861 & 14.337 & 10.552 & 35.920 & 3\% & 33,462 & 69.70\\
MS & 0.134 & 11.892 & 13.835 & 8.613 & 37.193 & 3\% & 59,597 & 124.64\\
\bottomrule
\end{tabular}

\caption{\label{tab:bench_stats} Percentage gap between the final integer objective and the reference optimum over 30 runs per solver (budget 2,000 evaluations), success rate (within 0.5\% of $f^* = 0.10892$) and expected running time to that target, in evaluations and seconds. Solvers are ordered by median gap.}
\end{table}

\begin{table}[htbp]
\centering
\scriptsize
\begin{tabular}{lrrrr}
\toprule
Solver vs PSO+PS & $U$ & $p$ & $p_{\mathrm{Holm}}$ & $\hat{P}$(PSO+PS better)\\
\midrule
PSO & 456 & 9.40e-01 & 9.40e-01 & 0.49\\
DPS & 152 & 1.00e-05 & 2.01e-05 & 0.83\\
SO & 115 & 7.07e-07 & 2.12e-06 & 0.87\\
GA & 0 & 2.66e-11 & 1.86e-10 & 1.00\\
SA & 1 & 2.94e-11 & 1.86e-10 & 1.00\\
GS & 0 & 2.66e-11 & 1.86e-10 & 1.00\\
MS & 0 & 2.66e-11 & 1.86e-10 & 1.00\\
\bottomrule
\end{tabular}

\caption{\label{tab:bench_tests} Pairwise two-sided Mann--Whitney tests of each solver against PSO+PS, with Holm-corrected $p$-values and the probability $\hat{P}$ that a run of PSO+PS ends with a lower objective (Vargha--Delaney $1-\hat{A}_{12}$, ties counted as one half).}
\end{table}

\begin{figure}[htbp]
  \centering
  \includegraphics[width=\linewidth]{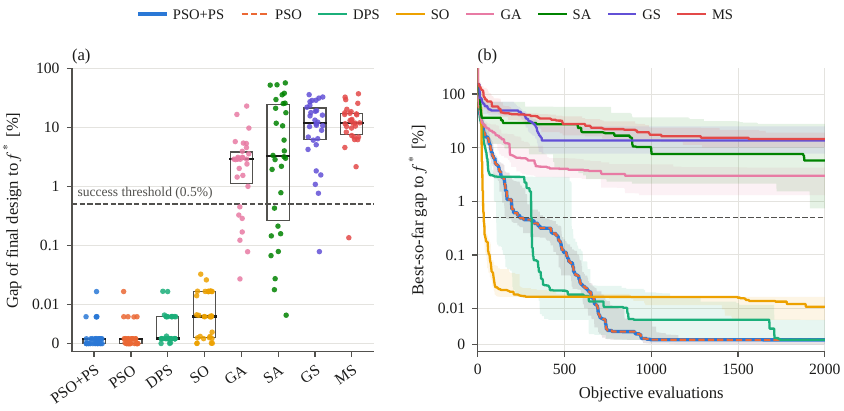}
  \caption{\label{fig:solver_comp} Benchmark over 30 runs per solver: (a) gap between the final integer design of each run and the integer optimum $f^*$ (dots: runs; boxes: interquartile range; bar: median; dashed line: success threshold of 0.5\%); (b) median best-so-far gap against objective evaluations, with interquartile bands. Both vertical axes are linear below 0.01\% and logarithmic above. PSO and PSO+PS share their first 80\% of evaluations, so their curves coincide.}
\end{figure}

\cref{tab:bench_stats} and \cref{fig:solver_comp} summarize the results. The Kruskal--Wallis test rejects equal performance across solvers ($H = 196.8$, $p = 5.6 \times 10^{-39}$). PSO+PS, PSO, DPS and SO reach the target in every run; GA (23\%), SA (30\%), GS (3\%) and MS (3\%) succeed in some runs. Holm-corrected Mann--Whitney tests against PSO+PS, the solver with the lowest median, find no significant difference for PSO and detect differences for DPS, SO, GA, SA, GS and MS (\cref{tab:bench_tests}), but the sizes of these differences vary greatly. For DPS and SO the median final objective lies within 0.007\% of $f^*$, as for PSO+PS, so the difference is far below any practical tolerance; for GA, SA, GS and MS the differences are large, with PSO+PS ending lower in almost every pairwise comparison. Among the solvers that reach the target in every run, the expected running time is lowest for PSO (1,518 evaluations, 2.9~s), followed by PSO+PS (1,575, 3.0~s); surrogate optimization is reliable but spends most of its time fitting the surrogate, so its ERT in seconds is the highest (24.0~s). PSO stops on its own stall criterion after a median of 1,501 evaluations, well within the budget. The gradient-based multistart methods perform worst, as expected on a stepped objective: their local solver stops on flat regions. Several integer designs are practically equivalent: 12 integer designs lie within 0.5\% of $f^*$, with no wind turbines and 286--297 PV modules, so solvers that succeed may end at slightly different designs with practically the same objective. We also ran PSO and GA on the continuous relaxation, rounding only the final point (relax-and-round with the same polish, 10 seeds each). On this problem the choice matters little: PSO reaches the target in 100\% of runs either way, and GA in 30\% with relax-and-round against 23\% with integer coding, because the optimum has no wind turbines and is flat in $n_s$.

One evaluation of the 8,760-hour simulation takes about 2.0~ms on the hardware of \cref{tab:protocol}, so a full PSO run takes a few seconds. The rule-based dispatch keeps evaluations fast; an inner optimization of hourly dispatch would multiply the cost of every evaluation. Based on these results we use integer-coded PSO for all subsequent sizing results (best of ten seeds with the integer polish): it reaches the target in every run with the lowest expected running time.

\subsection{Sizing results}

\subsubsection{\label{sec:strat} Strategy selection}

Sizing requires two operating decisions: when to replace the battery and whether surplus DG output may charge it.
\begin{itemize}
    \item The operator can replace the battery at fixed intervals set by its warranty or expected life (here every 15 years for LI), or when it has completed its rated number of charge--discharge cycles (5,475 for LI and 1,400 for LA). The second approach follows actual use and is the more realistic of the two.
    \item The DG has a minimum stable output. When the residual deficit is below this level, the DG still runs at its minimum and produces a surplus. This surplus can either be rectified and stored in the BS or sent to the dump load.
\end{itemize}

\begin{table}[htbp] 
\centering
\begin{tabular}{p{0.75cm} p{10cm}}
\toprule
\multicolumn{1}{c}{\textbf{Strategy}} & \multicolumn{1}{c}{\textbf{Description}}    \\
\midrule
1 & DE minimum-load surplus can charge LI; LI replaced every 15 y \\[1ex]
2 & DE surplus is dumped; LI replaced every 15 y \\[1ex]
3 & DE minimum-load surplus can charge LI; LI replaced every 5,475 cycles \\[1ex]
4 & DE surplus is dumped; LI replaced every 5,475 cycles \\
\bottomrule
\end{tabular}
\caption{Strategy selection for the LI+DE microgrid.}
\label{tab:strat_select} 
\end{table}

\cref{tab:strat} lists the best design found for each strategy (best of ten integer-coded PSO runs). Strategy 3 gives the lowest weighted objective (0.1089) and the lowest absolute LCOE (0.184~USD/kWh), and we use Strategy 3 in all subsequent results. Storing the generator's minimum-load surplus (Strategies 1 and 3) lowers dumped energy and fuel use relative to dumping it (Strategies 2 and 4). Cycle-based replacement (Strategies 3 and 4) costs less than a fixed 15-year schedule because the battery completes only about 195 equivalent full cycles a year, so its 5,475-cycle life outlasts the 25-year project.

The DE never runs in order to charge the battery. Under the load-following rule it starts only when renewable output and battery discharge cannot cover the load. When the residual deficit is smaller than the engine's minimum stable output ($0.3P_{DG,\;r}$), the engine still runs at that minimum, and the difference is surplus energy. Strategies 2 and 4 send this surplus to the dump load; Strategies 1 and 3 rectify it ($\eta_{rec} = 0.9$) and store it (round-trip efficiency 0.9). Both options burn the same fuel in that hour, so storage recovers energy that would otherwise be wasted. Battery wear from the extra throughput is accounted for through cycle-based replacement and capacity fade. To test whether the preference for Strategy 3 depends on cheap fuel or a long battery life, \cref{tab:strategy_stress} repeats the comparison of Strategies 3 and 4 at diesel prices up to 7.57~USD/gal (2.0~USD/L) and with the cycle life cut to 3,000. Strategy 3 retains the lower weighted objective in all six combinations, although Strategy 4 reaches a slightly lower absolute LCOE in the 3.20~USD/gal, 3,000-cycle case (by less than 0.001~USD/kWh). With the default 5,475-cycle life the battery completes 119--195 equivalent full cycles a year. With a 3,000-cycle life the optimum enlarges the battery to about 399~kWh, so that it completes about 120 cycles a year and lasts the 25-year project without replacement, and LCOE rises to 0.220--0.233~USD/kWh.

\begin{table}[htbp]
\centering
\small
\begin{tabular}{l l c c c c c c c}
\toprule
\textbf{Strategy} & \textbf{Soln. $[n_s, n_w, E_{b,init}]$} & \textbf{$J$} & \textbf{LCOE} & \textbf{Em.} & \textbf{DPSP} & \textbf{Dump} & \textbf{1-REF} & \textbf{LCOE [USD/kWh]} \\
\midrule
1 & [300, 0, 186.37] & 0.1400 & 0.4836 & 0.0447 & 0.0000 & 0.1429 & 0.0290 & 0.240 \\
2 & [304, 0, 189.05] & 0.1434 & 0.4887 & 0.0445 & 0.0000 & 0.1551 & 0.0285 & 0.243 \\
3 & [288, 0, 238.79] & 0.1089 & 0.3694 & 0.0442 & 0.0000 & 0.1012 & 0.0298 & 0.184 \\
4 & [292, 0, 262.09] & 0.1113 & 0.3808 & 0.0405 & 0.0000 & 0.1084 & 0.0270 & 0.189 \\
\bottomrule
\end{tabular}
\caption{\label{tab:strat} Best design for each strategy (16~kW DE, equal weights). LCOE and emissions are normalized by the diesel-only baseline (\cref{tab:baseline}).}
\end{table}

\begin{table}[htbp]
\centering
\small
\begin{tabular}{c c c c c c c c}
\toprule
\textbf{Diesel price} & \textbf{LI cycle life} & \multicolumn{2}{c}{\textbf{$J$}} & \multicolumn{2}{c}{\textbf{LCOE [USD/kWh]}} & \multicolumn{2}{c}{\textbf{Dump fraction}} \\
\textbf{[USD/gal]} & \textbf{[cycles]} & S3 & S4 & S3 & S4 & S3 & S4 \\
\midrule
3.20 & 5,475 & 0.1089 & 0.1113 & 0.184 & 0.189 & 0.101 & 0.108 \\
3.20 & 3,000 & 0.1146 & 0.1158 & 0.220 & 0.220 & 0.097 & 0.098 \\
5.00 & 5,475 & 0.0901 & 0.0919 & 0.208 & 0.212 & 0.103 & 0.104 \\
5.00 & 3,000 & 0.0914 & 0.0926 & 0.225 & 0.226 & 0.097 & 0.099 \\
7.57 & 5,475 & 0.0741 & 0.0755 & 0.228 & 0.233 & 0.100 & 0.099 \\
7.57 & 3,000 & 0.0742 & 0.0755 & 0.231 & 0.233 & 0.097 & 0.099 \\
\bottomrule
\end{tabular}
\caption{\label{tab:strategy_stress} Strategy 3 (DE surplus stored) vs Strategy 4 (DE surplus dumped) under higher diesel prices and shorter battery cycle life; best of five PSO runs per cell.}
\end{table}

\subsubsection{Technology selection\label{sec:tech_select}}
The MG designer also has to choose the battery and backup generator technologies. Four combinations were sized (\cref{tab:tech_select}). LI+DE gives the lowest weighted objective (0.1089), followed by LI+MT (0.1568), LA+DE (0.2566) and LA+MT (0.3456). The lead-acid battery needs a much larger nominal capacity (390~kWh with DE) because only 40\% of it is usable and its round-trip efficiency is 75\%, which raises its cost despite the lower price per kWh. LCOE and emissions are normalized with respect to a single-DG system of the same technology; the baseline values are listed in \cref{tab:baseline}.

\begin{table}[htbp]
\centering
\small
\begin{tabular}{l l c c c c c c}
\toprule
\textbf{System} & \textbf{Soln. $[n_s, n_w, E_{b,init}]$} & \textbf{$J$} & \textbf{LCOE} & \textbf{Em.} & \textbf{DPSP} & \textbf{Dump} & \textbf{1-REF} \\
\midrule
LI+DE & [288, 0, 238.79] & 0.1089 & 0.3694 & 0.0442 & 0.0000 & 0.1012 & 0.0298 \\
LI+MT & [276, 0, 208.77] & 0.1568 & 0.5908 & 0.0639 & 0.0000 & 0.0850 & 0.0444 \\
LA+DE & [296, 4, 389.98] & 0.2566 & 1.0405 & 0.0451 & 0.0000 & 0.1718 & 0.0257 \\
LA+MT & [229, 5, 293.55] & 0.3456 & 1.4268 & 0.1208 & 0.0000 & 0.1030 & 0.0776 \\
\bottomrule
\end{tabular}
\caption{\label{tab:tech_select} Best design for each technology combination (best of five to ten integer-coded PSO runs).}
\end{table}

\begin{table}[htbp]
\centering
\begin{tabular}{@{}lcc@{}}
\toprule
\textbf{Parameter} & DE system & MT system \\ 
\midrule
LCOE [USD/kWh] & 0.4968 & 0.3321 \\
Emissions [$kg$ of pollutants/y] & 50,189 & 48,562 \\
Emissions [$kg$ of pollutants/$kWh$] & 0.6774 & 0.6555 \\
\bottomrule
\end{tabular}
\caption{\label{tab:baseline} Baseline LCOE and emissions for the MG running on a single 16~kW fossil-fueled DG.}
\end{table}

Microturbines running on natural gas have lower specific emissions than diesel engines and, alone, a lower LCOE (\cref{tab:baseline}). Yet the MG with a DE backup has lower emissions and costs than the one with an MT backup, because the backup runs for only about 589~h per year: the MT's higher capital cost is not recovered by its fuel savings over so few operating hours. Lead-acid is costlier in both configurations: with DE backup its LCOE exceeds that of the diesel-only system.

\subsubsection{Optimal capacities for baseline LI+DE system}

We sized the LI+DE system with two sets of decision variables: (i) unit counts $[n_s, n_w, E_{b,\;init}]$ with integer $n_s$ and $n_w$, and (ii) total rated powers $[P_{PV,\;r,\;total}, P_{WT,\;r,\;total}, E_{b,\;init}]$ treated as continuous. Design (i) corresponds to [73.44~kW, 0.0~kW, 238.8~kWh] in rated terms and design (ii) to [73.84~kW, 0.00~kW, 235.3~kWh]. The two approaches give practically the same design, with objective values of 0.10892 for (i) and 0.10885 for (ii). The continuous approach (ii) gives a marginally lower objective because it is not restricted to whole PV modules and turbines, while a designer limited to specific products should use approach (i).

\begin{table}[htbp]
\centering
\begin{tabular}{p{5cm} p{4.3cm} p{4.3cm}}
\toprule
\multicolumn{1}{l}{\textbf{Parameter}} & \multicolumn{1}{l}{(i) unit counts} & \multicolumn{1}{l}{(ii) rated powers} \\
\midrule
$LCOE_{normalized}$ & 0.3694 & 0.3672 \\
LCOE [USD/kWh] & 0.1835 & 0.1824 \\
$Emissions_{normalized}$ & 0.0442 & 0.0429 \\
Annual emissions [$kg/y$] & 2,217 & 2,156 \\
DPSP & 0 & 0 \\
Dump & 0.1012 & 0.1053 \\
1-REF & 0.0298 & 0.0289 \\
No. of online DE hours [$h/y$] & 589 & 571 \\
No. of LI BS cycles [$/y$] & 195 & 198 \\
Optimal solution & $[n_s, n_w, E_{b,\; init}]$ = [288, 0, 238.79] & $[P_{PV}, P_{WT}, E_{b}]$ = [73.84, 0.00, 235.25] \\
Weighted objective $J$ & 0.1089 & 0.1089 \\
\bottomrule
\end{tabular}
\caption{\label{tab:LIDE} Sizing results for the baseline LI+DE system using two sets of decision variables.}
\end{table}

With either approach the hybrid microgrid improves substantially on the diesel-only baseline: LCOE falls by 63\% (from 0.497 to 0.184~USD/kWh), emissions by 96\%, and renewables supply 97.0\% of generation. The battery completes about 195--198 equivalent full cycles a year, so its 5,475-cycle life outlasts the 25-year project. The DE runs about 571--589~h a year, well within its 15,000-h life over 25 years.

\begin{figure}[htbp]
  \centering
  \includegraphics[width=\linewidth]{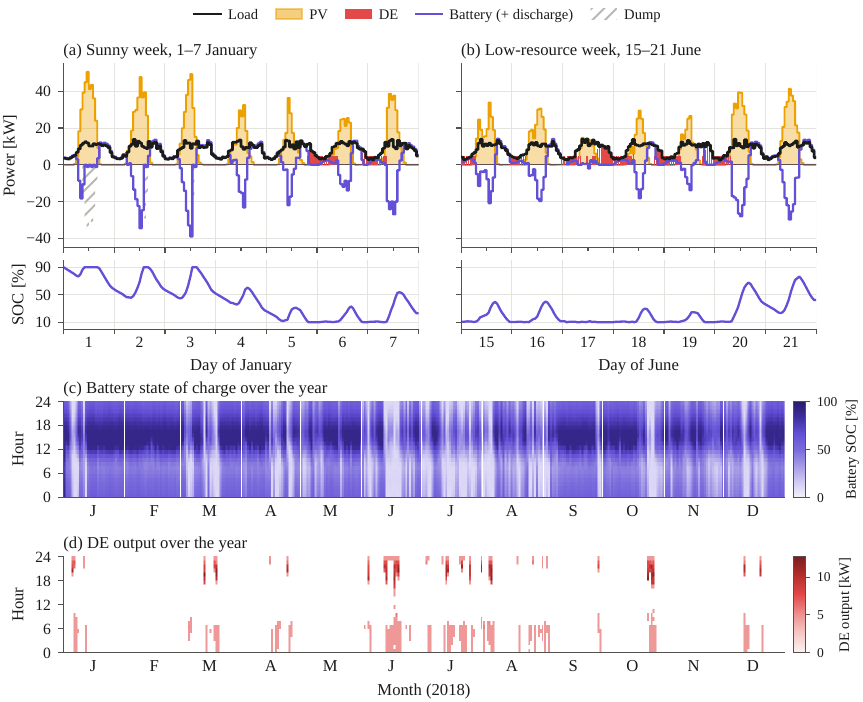}
  \caption{\label{fig:sizing_operation} Rule-based operation of the selected LI+DE design (16~kW DE): (a, b) hourly power flows and battery SOC in a sunny week (1--7 January) and a low-resource week (15--21 June); battery power is positive when discharging and dumped power is drawn below zero; (c) battery SOC and (d) DE output for every hour of 2018 (blank: DE off).}
\end{figure}

The renewable capacity of design (i) is 73.4~kW, 8.7 times the average load of 8.46~kW and 4.9 times the annual peak of 15.0~kW, and the battery holds 1.18 days of average consumption (203~kWh/day). This capacity is needed to supply 97\% of the energy from a single variable source, with 10\% of generated energy dumped. \cref{fig:sizing_operation} shows the resulting rule-based operation. In the first week of January the battery fills to its upper limit on the first three days, when surplus PV is dumped, and the DE starts on the last three nights as the battery runs down. In the low-resource week of 15--21 June the battery stays near its lower limit for five days and the DE runs on six of the seven days, for 19~h on 17 June. Over the year the DE runs for 589~h on 83 days, mostly in June--August, which account for 55\% of its output; 93\% of its energy is delivered between 17:00 and 07:00, and in 80\% of its operating hours it runs at its minimum stable output of 4.8~kW. The optimum installs no wind turbines. With the scaled station data, a 3.5~kW turbine at the 14.5~m hub height has a capacity factor of 14\%, close to the 15\% of a PV module, but wind costs 1,500~USD/kW installed against 1,210~USD/kW for PV, so PV supplies the renewable energy more cheaply; global averages of installed cost were used for both, and small turbines usually cost more per kW than these averages, which would further favor PV. \cref{sec:wind_role} examines why, and at what wind resource and turbine cost wind would enter the mix.

\subsubsection{Role of wind turbines\label{sec:wind_role}}
The optimal design installs no wind turbines. We examine the cost of adding wind and the resource and cost conditions under which wind would become economical.

\begin{figure}[htbp]
  \centering
  \includegraphics[width=\linewidth]{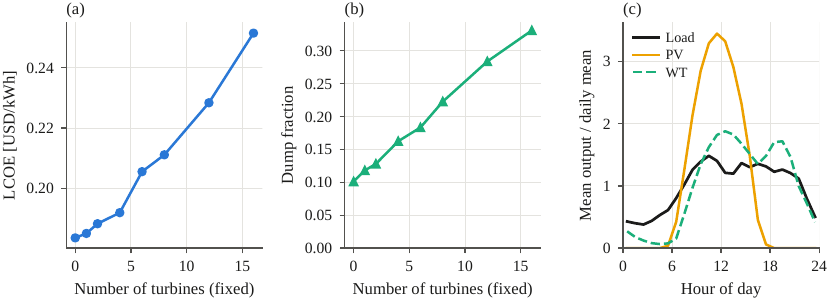}
  \caption{\label{fig:wind_role} Role of wind turbines: (a) LCOE and (b) dump fraction when the number of turbines is fixed and the PV array and battery are re-optimized (\cref{tab:wind_forced}); (c) mean diurnal profiles of load, PV output and wind output, each divided by its daily mean, for the design with four turbines.}
\end{figure}

\textit{Cost of including wind.} We fixed the number of turbines and re-optimized the PV array and battery (best of five PSO runs with a local polish of $E_{b,\;init}$; \cref{tab:wind_forced} and \cref{fig:wind_role}a--b). With four turbines, LCOE rises by 4.6\% (from 0.184 to 0.192~USD/kWh), the weighted objective rises by 14\%, and annual emissions are essentially unchanged. Battery capacity falls by 13\% and the PV array loses 28 modules. The penalty grows steadily with the number of turbines.

\textit{Limited complementarity between wind and solar.} At this site the wind blows mostly in daylight hours, at the same time as the PV output (\cref{fig:wind_role}c): hourly PV and wind output are positively correlated ($r = 0.30$), and 53\% of the wind energy arrives in hours when PV alone already exceeds the load, so much of it is dumped. Wind does continue into the late afternoon and evening (47\% of wind energy falls after 15:00 local time), which is why it lets the battery shrink. Seasonally, wind is about 2.8 times stronger from August to November than in the rest of the year, but PV output does not drop in those months, so the two resources do not complement each other across seasons either (monthly $r = 0.27$). Under the turbine power curve a 3.5~kW turbine has a capacity factor of 14\%, close to the 15\% of a PV module, at a higher installed cost per kW.

\textit{Conditions favoring wind.} We swept the wind resource and the installed turbine cost (best of three to five PSO runs per point). Under the equal-weight objective, turbines enter only when the wind input is scaled by 1.7 (a hub-height mean of about 8.0~m/s and a turbine capacity factor of 42\%), and they do not enter at any turbine cost down to 300~USD/kW, because the extra midday energy raises the dump fraction, which the objective penalizes. Under a cost-only objective ($w_1 = 1$), wind becomes competitive at smaller changes in resource or cost: turbines enter at an installed cost of 1,050~USD/kW or below (30\% below our assumption), or when the wind input is 20\% stronger (hub-height mean 5.6~m/s). At 600~USD/kW the cost-optimal design has 5 turbines, a 20\% smaller battery and a 3\% lower LCOE than the PV-only design. Wind can therefore become economical at sites with a stronger or more nocturnal wind resource, or where small turbines cost substantially less than assumed here; at Timbila, with the current data and costs, PV with storage is the better choice.

\begin{table}[htbp]
\centering
\small
\begin{tabular}{c l c c c c c c}
\toprule
\textbf{$n_w$ (fixed)} & \textbf{$[n_s, n_w, E_b]$} & \textbf{$J$} & \textbf{LCOE [USD/kWh]} & \textbf{Em. [kg/y]} & \textbf{Dump} & \textbf{REF [\%]} & \textbf{DE [h/y]} \\
\midrule
0 & [288, 0, 239] & 0.1089 & 0.184 & 2,217 & 0.101 & 97.0 & 589 \\
1 & [282, 1, 231] & 0.1120 & 0.185 & 2,101 & 0.118 & 97.2 & 558 \\
2 & [273, 2, 229] & 0.1155 & 0.188 & 2,133 & 0.128 & 97.2 & 562 \\
4 & [260, 4, 208] & 0.1241 & 0.192 & 2,207 & 0.162 & 97.2 & 574 \\
8 & [237, 8, 210] & 0.1437 & 0.211 & 2,233 & 0.223 & 97.4 & 580 \\
16 & [198, 16, 215] & 0.1810 & 0.251 & 2,265 & 0.331 & 97.7 & 582 \\
\bottomrule
\end{tabular}
\caption{\label{tab:wind_forced} Best design with a fixed number of wind turbines (16~kW DE, equal weights); PV array and battery re-optimized for each $n_w$.}
\end{table}

\subsection{Pareto approximation\label{sec:pareto_opt}}

The weighted-sum problem in Eq.~(\ref{eq:sizing}) has a single minimum value, but the five-objective problem behind it has no single optimum: along the Pareto front, any improvement in one objective worsens at least one other. A design belongs to the Pareto set if no other feasible design is at least as good in every objective and strictly better in at least one. We approximate this set with the pattern-search multiobjective solver \texttt{paretosearch} in MATLAB's Global Optimization Toolbox (200 points, integer-coded unit counts, 16~kW DE). The points are mutually non-dominated: they represent different trade-offs, not equally good designs, and choosing among them requires stated preferences or a selection rule (\cref{sec:pareto_select}).

\begin{figure}[htbp]
  \centering
  \includegraphics[width=\linewidth]{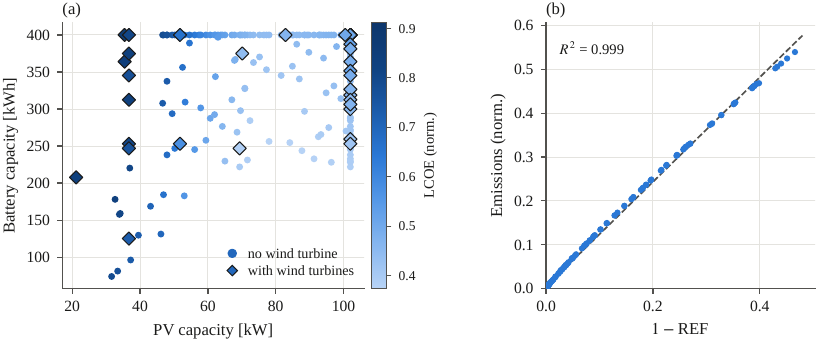}
  \caption{\label{fig:pareto_inputs} The 16~kW Pareto approximation (\texttt{paretosearch}, five objectives): (a) battery vs PV capacity of the 200 designs, colored by normalized LCOE; diamonds mark the 39 designs with wind turbines; (b) normalized emissions vs $(1-REF)$, with a linear fit.}
\end{figure}

\cref{fig:pareto_inputs}a shows the designs on the front. PV capacity ranges from 21 to 102~kW and battery capacity from 74 to 400~kWh; 79 of the 200 points sit at the 400~kWh upper bound, at the low-emission end of the front. Of the 200 designs, 161 have no wind turbines; the rest have at most 18~kW of wind capacity, mainly in low-dump designs with small PV arrays. Normalized emissions and $(1-REF)$ are almost perfectly correlated along the front ($R^2 = 0.999$, \cref{fig:pareto_inputs}b), because the DE is the only source of both; one of the two objectives is therefore redundant for this system. With a 16~kW DE, DPSP is zero everywhere on the front.

\begin{figure}[htbp]
  \centering
  \includegraphics[width=\linewidth]{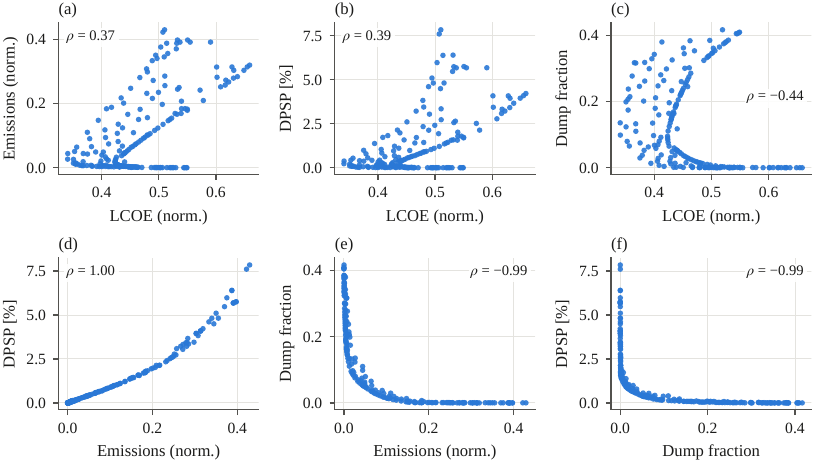}
  \caption{\label{fig:pareto_obj} Pairwise relationships between the objectives on the four-objective Pareto approximation with a 7~kW DE (\texttt{paretosearch}, 200 points); $\rho$: Spearman rank correlation.}
\end{figure}

To expose the reliability dimension we also computed a four-objective front (LCOE, emissions, DPSP and dump) with a 7~kW DE and rated-power decision variables (\cref{fig:pareto_obj}). Along this front, dumped energy trades off almost one-for-one against emissions and DPSP (Spearman $\rho = -0.99$ and $-0.99$): avoiding dumped energy requires smaller renewable arrays, which leaves more load to the small DE and, when the DE is at its rating, unserved. Emissions and DPSP rise together ($\rho = 1.00$), so for this system the cleanest designs are also the most reliable. LCOE is only weakly related to the other objectives along the front ($|\rho| \leq 0.44$): the cost-optimal region is flat, and moderate changes in emissions or dump cost little.

\subsection{Selecting designs from the Pareto set\label{sec:pareto_select}}
\begin{figure}[htbp]
  \centering
  \includegraphics[width=\linewidth]{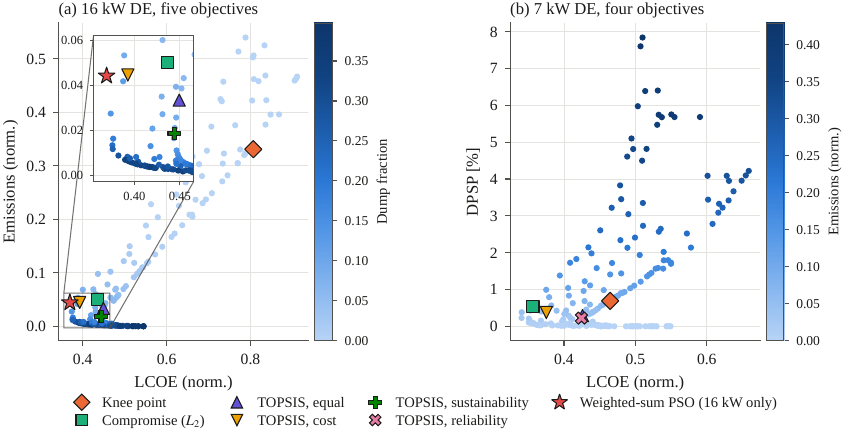}
  \caption{\label{fig:pareto_reps} Pareto approximations and the representative designs of \cref{tab:pareto_reps}: (a) 16~kW DE, normalized LCOE vs emissions colored by dump fraction, with the weighted-sum PSO design; the inset enlarges the low-cost region; (b) 7~kW DE, normalized LCOE vs DPSP colored by normalized emissions.}
\end{figure}

Choosing a design from a Pareto set requires explicit preferences or a selection rule. We extract four kinds of representative design from each front. All objectives are first min--max normalized over the front, and objectives that are constant on the front (DPSP with the 16~kW DE) are dropped.
\begin{itemize}
\item \textit{Extreme solutions}: the best design for each single objective.
\item \textit{Knee point}: the design with the largest distance below the hyperplane through the extreme solutions, where a small gain in one objective costs a large loss in another \cite{deb2011knee}.
\item \textit{Compromise solution}: the design closest to the ideal point in the $L_2$ norm.
\item \textit{TOPSIS} \cite{hwang1981topsis}: the design with the highest relative closeness to the ideal and distance from the anti-ideal solution, under three weightings: equal; cost priority (four times the weight on LCOE); and sustainability priority (2.5 times the weight on emissions and $(1-REF)$) for the 16~kW front, or reliability priority (three times the weight on DPSP) for the 7~kW front.
\end{itemize}
Each representative is rounded to whole PV modules and turbines and re-evaluated exactly (\cref{tab:pareto_reps}, \cref{fig:pareto_reps}).

On the 16~kW front the cost-priority TOPSIS choice, [284, 0, 284], has an LCOE of 0.195~USD/kWh, close to the equal-weight design, while the compromise solution, [278, 0, 328], uses a larger battery at an LCOE of 0.217~USD/kWh. Equal-weight and sustainability-priority TOPSIS choose the largest battery allowed (400~kWh), reducing emissions to 1,675 and 927~kg/y at LCOEs of 0.223 and 0.220~USD/kWh, respectively. The knee point, [144, 3, 400], sits where dumped energy falls to almost zero; its LCOE is 0.401~USD/kWh and its emissions are 16,602~kg/y, so a designer who does not value dump reduction highly would not choose it. As a consistency check, none of the 190 non-dominated front points dominates the equal-weight PSO design, and its equal-weight score (0.1089) is lower than that of every front point (best 0.1104). The \texttt{paretosearch} approximation therefore does not sample the front densely near this design, and we add the weighted-sum design to the candidate set. \cref{sec:mc} re-ranks all candidates by their performance under uncertainty.

\begin{table}[htbp]
\centering
\scriptsize
\setlength{\tabcolsep}{3pt}
\begin{tabular}{l l l c c c c c c}
\toprule
\textbf{DE} & \textbf{Selection rule} & \textbf{$[n_s, n_w, E_b]$} & \textbf{LCOE} & \textbf{Em.} & \textbf{DPSP} & \textbf{Dump} & \textbf{REF} & \textbf{$J$} \\
 & & & [USD/kWh] & [kg/y] & [\%] & & [\%] & \\
\midrule
16~kW & Weighted-sum PSO & [288, 0, 239] & 0.184 & 2,217 & 0.00 & 0.101 & 97.0 & 0.1089 \\
16~kW & Min LCOE & [306, 0, 256] & 0.186 & 1,375 & 0.00 & 0.136 & 98.2 & 0.1109 \\
16~kW & Min emissions & [400, 5, 400] & 0.271 & 0 & 0.00 & 0.391 & 100.0 & 0.1871 \\
16~kW & Min dump & [124, 0, 74] & 0.392 & 27,073 & 0.00 & 0.000 & 53.4 & 0.3587 \\
16~kW & Knee & [144, 3, 400] & 0.401 & 16,602 & 0.00 & 0.000 & 73.0 & 0.2816 \\
16~kW & Compromise ($L_2$) & [278, 0, 328] & 0.217 & 2,508 & 0.00 & 0.066 & 96.5 & 0.1174 \\
16~kW & TOPSIS, equal & [288, 0, 400] & 0.223 & 1,675 & 0.00 & 0.073 & 97.7 & 0.1156 \\
16~kW & TOPSIS, cost & [284, 0, 284] & 0.195 & 2,236 & 0.00 & 0.084 & 97.0 & 0.1104 \\
16~kW & TOPSIS, sustainability & [303, 0, 400] & 0.220 & 927 & 0.00 & 0.100 & 98.8 & 0.1149 \\
7~kW & Min LCOE & [303, 0, 234] & 0.169 & 1,345 & 0.23 & 0.135 & 98.3 & 0.1044 \\
7~kW & Min emissions & [392, 7, 400] & 0.274 & 0 & 0.00 & 0.411 & 100.0 & 0.1923 \\
7~kW & Min dump & [125, 3, 400] & 0.330 & 16,520 & 4.36 & 0.000 & 70.8 & 0.2659 \\
7~kW & Knee & [253, 0, 400] & 0.231 & 3,983 & 0.69 & 0.025 & 94.1 & 0.1269 \\
7~kW & Compromise ($L_2$) & [267, 0, 222] & 0.177 & 3,185 & 0.54 & 0.066 & 95.5 & 0.1074 \\
7~kW & TOPSIS, equal & [283, 0, 400] & 0.212 & 1,648 & 0.29 & 0.065 & 97.7 & 0.1099 \\
7~kW & TOPSIS, cost & [279, 0, 284] & 0.186 & 2,123 & 0.37 & 0.075 & 97.0 & 0.1052 \\
7~kW & TOPSIS, reliability & [290, 0, 400] & 0.211 & 1,281 & 0.22 & 0.077 & 98.3 & 0.1092 \\
\bottomrule
\end{tabular}
\caption{\label{tab:pareto_reps} Representative designs selected from the 16~kW (five objectives) and 7~kW (four objectives) Pareto approximations, after rounding to whole units and exact re-evaluation. $J$: equal-weight objective. Max REF coincides with min emissions and is not listed.}
\end{table}

\subsection{Cost--reliability trade-off with the DE rating as a decision variable\label{sec:pareto_rel}}
With a DE rated above the peak load, DPSP is zero for every design and the reliability objective never trades off against the others. We expose this trade-off in two ways. First, we add the DE rating (0--20~kW) to the decision variables, with its capital, replacement, maintenance and fuel costs scaling with rating. Under equal weights, all 10 PSO runs converge to a design without a generator (0~kW): the design [245, 0, 181] leaves 11.0\% of the load unserved at an LCOE of 0.129~USD/kWh. The equal-weight objective thus permits substantial unmet load. An explicit reliability constraint is needed to enforce a community's service requirement (see also \cref{sec:sizing_sensitivity}).

\begin{figure}[htbp]
  \centering
  \includegraphics[width=\linewidth]{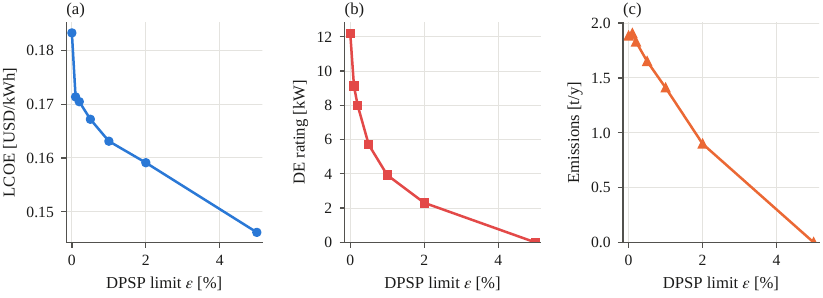}
  \caption{\label{fig:dg_front} Cost--reliability front from the $\varepsilon$-constraint problem with the DE rating as a decision variable: (a) LCOE, (b) DE rating and (c) annual emissions vs the DPSP limit $\varepsilon$. The PV array (71--74~kW) and battery (222--262~kWh) change little along the front.}
\end{figure}

Second, we solve an $\varepsilon$-constraint problem: minimize the mean of the other four normalized objectives subject to DPSP $\leq \varepsilon$, with the DE rating free, for $\varepsilon$ from 0 to 5\% (best of five PSO runs each). \cref{fig:dg_front} shows the resulting cost--reliability front. The selected zero-unmet-load design uses a 12.2~kW DE, below the 15.0~kW hourly peak because the battery covers part of the peak, at an LCOE of 0.183~USD/kWh, about the same as the equal-weight design with the default 16~kW DE (0.184~USD/kWh). Allowing 1\% unmet load reduces the DE to 3.9~kW and LCOE to 0.163~USD/kWh; at 5\% the generator disappears and LCOE falls to 0.146~USD/kWh. The PV array and battery hardly change along the front (71--74~kW and 222--262~kWh): the reliability target mainly affects the size of the backup generator.

\subsection{Break-even distance analysis}

The break-even distance (BED) is the distance from the existing grid beyond which the stand-alone MG becomes cheaper than extending the grid \cite{kaabeche2011a}:
\begin{align*}
    BED \; [km] & = \frac{TAC - LCOE_{grid} \cdot \Sigma_{t=1}^{t=8760}\; P_{load}(t)\cdot \Delta t}{C_{ext} \cdot CRF}
\end{align*}
where $LCOE_{grid}$ is the levelized cost of grid electricity and $C_{ext}$ the cost of line extension. For the selected design, $TAC = 13,597$~USD/y and $CRF = 0.0582$. With $LCOE_{grid} = 0.125$~USD/kWh \cite{LCOEgrid} and $C_{ext} = 157,470$~USD/km \cite{Cext} for Kenya, buying the community's 74,086~kWh from the grid would cost 9,261~USD/y before any extension cost. The MG's annualized cost is 4,336~USD/y higher, but each kilometer of line extension costs 9,165~USD/y when annualized, so the break-even distance is only 0.47~km: farther from the grid than this, the MG is the cheaper option. Timbila lies about 111~km from Voi, the nearest large town, and 305~km from Nairobi.

The MG's LCOE of 0.184~USD/kWh is within the 0.12--0.20~USD/kWh residential grid tariff in Kenya \cite{LCOEgrid}, but this tariff excludes connection fees of 136--400~USD per household, which can be prohibitive for rural customers \cite{quak_CBA, golumbeanu2013}. Grid customers in many peri-urban areas also face frequent outages \cite{iea_africa2019}. Households without electricity pay far more for kerosene lighting (1.00--2.50~USD/kWh equivalent) \cite{tracy2012} or small diesel generators (0.80--1.50~USD/kWh) \cite{quak_CBA}. This comparison rests on the model's cost assumptions (IRENA 2018 installed costs and a real discount rate of 3.1\%); \cref{sec:sizing_sensitivity} shows how LCOE responds to them.

\section{Sensitivity analysis of sizing optimization\label{sec:sizing_sensitivity}}

\begin{figure}[htbp]
  \centering
  \includegraphics[width=0.98\linewidth]{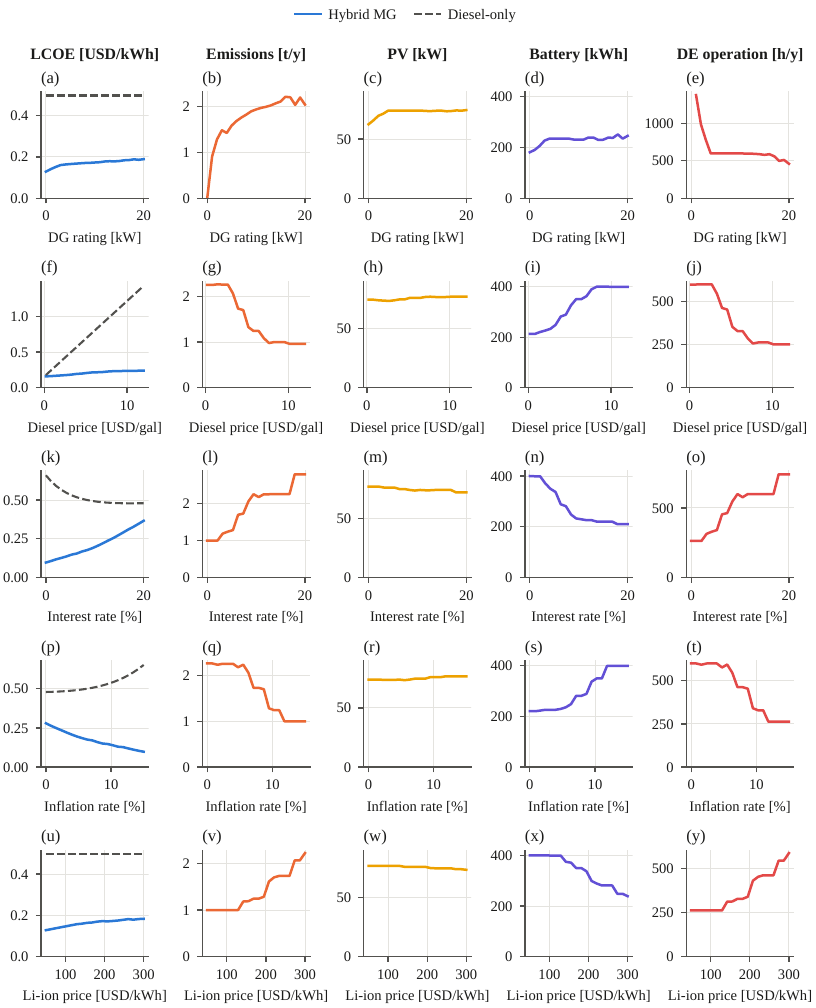}
  \caption{\label{fig:sensitivity} One-at-a-time sensitivity of the sized design to the DG rating (a--e), diesel price (f--j), nominal interest rate (k--o), inflation rate (p--t) and Li-ion battery price (u--y). Columns: absolute LCOE of the hybrid MG and of the diesel-only baseline; annual emissions; rated PV capacity; battery capacity; annual DE operating hours (not shown at 0~kW, where no generator is installed). No design installs wind turbines. Each point is the best of three integer-coded PSO runs.}
\end{figure}

We assess the sensitivity of the design by varying one parameter at a time and re-optimizing the sizing problem at each value; all other parameters retain their default values in \cref{tab:default}. Each point is the best of three integer-coded PSO runs. The upper bound on battery capacity (400~kWh) binds only at the lowest interest rates, the highest inflation rates and the lowest battery prices; there the optimum would install more storage. Because LCOE is normalized by a diesel-only baseline that itself depends on fuel price and financial parameters (\cref{app:costs}), \cref{fig:sensitivity} reports absolute LCOE together with that of the diesel-only baseline, and annual emissions, alongside the PV capacity, battery capacity and DE operating hours.

\begin{table}[htbp]
\centering
\begin{tabular}{@{}lc@{}}
\toprule
\textbf{Parameter}                 & \textbf{Default value}                        \\ \midrule
DG power rating [$kW$]         & 16                                            \\
Diesel fuel price [USD/US gal] & 3.20                                          \\
Nominal interest rate (\%)         & 9                                             \\
Inflation rate (\%)                & 5.70                                           \\
Price of BS (USD/kWh)               & 300  \\
{[}$w_1$ $w_2$ $w_3$ $w_4$ $w_5${]}  &  [0.2 0.2 0.2 0.2 0.2] \\ 
\bottomrule
\end{tabular}
\caption{\label{tab:default} Default parameter values used for sensitivity analysis.}
\end{table}

\subsection{Varying DG power rating}
With the default 16~kW DG, DPSP is zero for every design because the DG exceeds the annual peak load. We varied the DG rating from 0 to 20~kW, keeping the normalization baseline at 16~kW (\cref{fig:sensitivity}a--e). DPSP is zero for ratings of 13~kW and above, and stays below 1\% down to 4~kW; an 8~kW DG gives DPSP = 0.0020. LCOE is 0.170~USD/kWh with an 8~kW DG, 0.184 with 16~kW and 0.189 with 20~kW; larger generators add capital and fixed O\&M costs. Without any DG, 11.0\% of the load is lost, yet the equal-weight objective ($J$ = 0.0822) is the lowest of all ratings: with equal weights, a 10\% loss of load adds only 0.02 to $J$, so the scalarized objective undervalues reliability for an off-grid community. \cref{sec:pareto_rel} therefore treats reliability as an explicit constraint.

\subsection{Varying fuel price}
The diesel price was varied from 0.2 to 12~USD/gal, a range based on reported global prices \cite{diesel_price} (\cref{fig:sensitivity}f--j). The absolute LCOE of the hybrid MG rises from 0.161 to 0.241~USD/kWh, while the normalized LCOE falls, because the diesel-only baseline rises much faster (from 0.180 to 1.427~USD/kWh). Across the price range the battery changes from 213 to 398~kWh, the PV array from 74 to 77~kW and DE operation from 599 to 252~h per year. At the 2.0~USD/L (7.57~USD/gal) level typical of remote sites the hybrid system's advantage over diesel-only supply grows further.

\subsection{\label{sec:varyInterest} Varying nominal interest or discount rate}
The nominal interest rate sets the opportunity cost of capital and hence the trade-off between capital-intensive renewables and storage and fuel-intensive backup generation. We varied it from 0 to 20\%, a range based on reported central bank rates \cite{interest_rates} (\cref{fig:sensitivity}k--o). With inflation fixed at 5.7\%, the real discount rate $r = (i - f)/(1 + f)$ ranges from $-5.4\%$ to $13.5\%$. Higher rates penalize upfront investment: from 0 to 20\% the battery changes from 400 to 210~kWh, the PV array from 77 to 72~kW, DE operation from 263 to 742~h per year, and LCOE from 0.097 to 0.366~USD/kWh. Interest rates in many SSA countries sit in the upper half of this range, which makes concessional finance important for the affordability of off-grid renewables.

\subsection{Varying inflation rate}
Inflation was varied from 0 to 15\%, a range based on reported international values \cite{inflation_rate} (\cref{fig:sensitivity}p--t). At a fixed nominal interest rate of 9\%, higher inflation lowers the real discount rate, so the effect mirrors that of the interest rate: as inflation rises from 0 to 15\%, LCOE changes from 0.280 to 0.098~USD/kWh and the battery from 220 to 399~kWh.

\subsection{Varying price of battery energy storage (BS)}
Li-ion prices have fallen steeply and are projected to keep falling \cite{bnef}, so we varied the battery price from 50 to 300~USD/kWh (\cref{fig:sensitivity}u--y). At 50~USD/kWh the optimal battery is 400~kWh, against 239~kWh at 300~USD/kWh; DE operation is 263 vs 589~h per year and LCOE 0.128 vs 0.184~USD/kWh.

\subsection{Varying the wind-speed correction factor\label{sec:windsens}}
The wind input rests on station data scaled by a factor of 3.70 (\cref{app:climate}). A comparison with NASA POWER reanalysis suggests that the resulting hub-height wind may be about 25\% too high, so we re-optimized the design with the wind input scaled by 0.75, 1.00 and 1.25, i.e.\ with correction factors of 2.78, 3.70 and 4.63. All three runs return the same design, [288, 0, 239] with no wind turbines, at an LCOE of 0.184~USD/kWh, so the uncertain wind level does not affect the base result, and the hybrid system with renewables, Li-ion storage and DE backup beats diesel-only supply across this range.

\section{\label{sec:dispatch_opt} Optimal economic and environmental dispatch results}

The sizing stage (Stage~1) scores each design under the rule-based, priority-ordered controller of \cref{sec:disp_strat}, which is fast and non-anticipative but does not minimize cost or emissions over time. Stage~2 asks how much optimized day-ahead scheduling changes the operation of a given design. For a fixed design it chooses the hourly DE output $P_{DE}(t)$ and battery power $P_{BS}(t)$ over the next 24~h to minimize the equally weighted sum of four normalized terms, daily COE, emissions, dumped energy and $(1-REF)$ (Eq.~\ref{eq:dispatch}). DPSP enters as a constraint ($DPSP \leq 1\%$) together with the SOC limits (10--90\%), the DE output limits and the battery power rating; lost load carries no cost in this objective, so the optimizer may use the permitted 1\% where that lowers the other terms. The daily cost includes battery wear at 0.068~USD per kWh discharged (the replacement cost of 300~USD/kWh divided by the lifetime throughput of 5,475 cycles over 80\% of capacity), so the optimizer does not cycle the battery without need. The weighted sum is again an a priori scalarization (\cref{sec:opt_prob}). The dispatch assumes perfect foresight of load and weather over the day, which makes its results optimistic (\cref{sec:limitations}). We solve it in two steps: particle swarm optimization on a penalty form of the problem (five seeds, swarm size 100, 600 iterations) followed by \texttt{fmincon} with the SQP algorithm, started from each PSO result, from a flat schedule and from a load-following schedule that respects the SOC limits, enforcing the constraints exactly; the best feasible solution is kept. Each case takes about 5~s.

\cref{tab:DG_size} compares three configurations on the representative day (1 January 2018): (i) the equal-weight design with a 16~kW DE, (ii) the same design with an 8~kW DE, and (iii) the design re-sized for an 8~kW DE ([290, 0, 235]). The representative day is sunny (irradiation 136\% of the annual daily mean), and all three configurations operate without the DE, meeting demand from PV and storage with up to 1\% unmet load: emissions are zero, the daily COE (variable and daily fixed operating costs, including battery wear, divided by the day's demand) is 0.057--0.059~USD/kWh, and 52\% of the generated energy is dumped because the battery reaches its upper SOC limit by midday. The smaller DE lowers the fixed cost of the backup, so cases (ii) and (iii) have a lower COE than case (i); the small difference between (ii) and (iii), whose designs are almost identical, reflects the tolerance of the dispatch solver.

\begin{figure}[htbp]
  \centering
  \includegraphics[width=\linewidth]{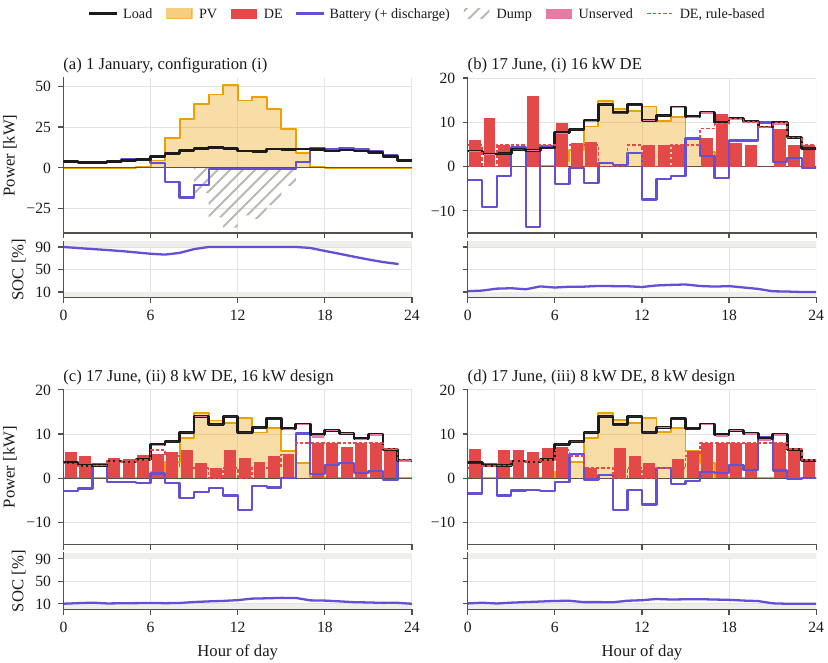}
  \caption{\label{fig:dispatch_P_base} Optimized day-ahead schedules: (a) 1 January, configuration (i), with which configurations (ii) and (iii) practically coincide; (b--d) 17 June, configurations (i)--(iii) of \cref{tab:DG_size}, with the DE output of the rule-based schedule (dashed) for comparison. Power flows on top (battery power positive when discharging; dumped power below zero; unserved load stacked on the load line) and battery SOC below.}
\end{figure}

Because the DE stays off on the representative day, we repeat the comparison on 17 June 2018, the day on which the equal-weight design burns the most diesel under rule-based operation (19~h and 122~kWh of DE output). It is the calmest day of the year and the fourth-least sunny (irradiation 35\% of the annual daily mean), and each configuration starts it with the battery almost empty, at the SOC that its own rule-based simulation reaches at midnight (10--11\%). To compare the two controllers directly, \cref{tab:DG_size} also scores the rule-based schedule of that day with the dispatch cost model. With the 16~kW DE, optimized dispatch changes little: it runs the DE for 17 instead of 19~h, burns 6\% less fuel (51.6 vs 54.9~L) and lowers the daily COE from 0.290 to 0.289~USD/kWh, partly by using the 1\% unmet-load allowance that the rule-based controller does not use. For the selected design, rule-based operation thus costs almost the same as optimized scheduling on the day with the most diesel use. With an 8~kW DE the difference lies in reliability rather than cost. The rule-based controller depletes the usable battery energy before the evening, when the load exceeds the DE rating, and leaves 5.2\% of the day's load unserved; the optimized schedule runs the DE early in the day to charge the battery and keeps unmet load within the 1\% limit, at a 7--8\% higher daily COE (0.243 vs 0.226~USD/kWh) because it serves more load. On this day the 8~kW configurations cost 16\% less per kWh than the 16~kW one, with the same emissions (80--81~kg), because the smaller engine burns less no-load fuel. \cref{fig:dispatch_P_base} shows the schedules.

\begin{table}[htbp]
\centering
\small
\begin{tabular}{l c c c c c c c c}
\toprule
\textbf{Case} & \textbf{$J$} & \textbf{COE} & \textbf{COE} & \textbf{Em.} & \textbf{DPSP} & \textbf{Dump} & \textbf{REF} & \textbf{DE} \\
 & & [USD/kWh] & (norm.) & [kg] & [\%] & & [\%] & [h] \\
\midrule
\multicolumn{9}{l}{\textit{1 January (representative day), optimized dispatch}} \\
(i) 16~kW DE & 0.1694 & 0.059 & 0.1592 & 0.0 & 0.00 & 0.518 & 100.0 & 0 \\
(ii) 8~kW DE, design sized for 16~kW & 0.1677 & 0.057 & 0.1526 & 0.0 & 1.00 & 0.518 & 100.0 & 0 \\
(iii) 8~kW DE, design sized for 8~kW & 0.1686 & 0.057 & 0.1524 & 0.0 & 1.00 & 0.522 & 100.0 & 0 \\
\midrule
\multicolumn{9}{l}{\textit{17 June (low-resource day), optimized dispatch}} \\
(i) 16~kW DE & 0.4698 & 0.289 & 0.7810 & 81.1 & 1.00 & 0.000 & 48.0 & 17 \\
(ii) 8~kW DE, design sized for 16~kW & 0.4392 & 0.243 & 0.6568 & 81.3 & 0.98 & 0.000 & 48.0 & 21 \\
(iii) 8~kW DE, design sized for 8~kW & 0.4357 & 0.243 & 0.6562 & 80.1 & 1.00 & 0.000 & 48.5 & 19 \\
\midrule
\multicolumn{9}{l}{\textit{17 June, rule-based operation of the sizing stage}} \\
(i) 16~kW DE & 0.4753 & 0.290 & 0.7844 & 82.9 & 0.00 & 0.000 & 47.5 & 19 \\
(ii) 8~kW DE, design sized for 16~kW & 0.4075 & 0.226 & 0.6116 & 73.4 & 5.20 & 0.000 & 50.5 & 22 \\
(iii) 8~kW DE, design sized for 8~kW & 0.4044 & 0.226 & 0.6099 & 72.6 & 5.19 & 0.000 & 51.0 & 22 \\
\bottomrule
\end{tabular}
\caption{\label{tab:DG_size} Day-ahead dispatch of three DE and design configurations on the representative day (1 January 2018) and on the low-resource day (17 June 2018), and the rule-based schedule of 17 June scored with the same cost model. $J$ is the equally weighted sum of normalized COE, emissions, dump fraction and $(1-REF)$; DPSP is constrained to at most 1\% in the optimized dispatch but not in the rule-based schedule, so $J$ does not penalize the unmet load in rule-based cases (ii) and (iii).}
\end{table}

\section{Robustness to resource and demand uncertainty}

\subsection{Monte Carlo analysis of resource and load uncertainty\label{sec:mc}}
The sizing results rest on one year of weather and one load profile. To quantify how annual performance varies, we simulate each candidate design over 500 synthetic years. Each scenario is built in three steps. (1) Day-block bootstrap: for each calendar day we draw a donor day from the same month of 2018 and copy its 24-hour irradiance, temperature and wind series, which preserves diurnal shapes, cross-correlation between the variables and seasonality. (2) Interannual variability: the whole year's irradiance is scaled by $a_I \sim \mathcal{N}(1, 0.05^2)$ and wind speed by $a_W \sim \mathcal{N}(1, 0.10^2)$. (3) Load noise: hourly load is multiplied by $(1 + \epsilon_t)$, where $\epsilon_t$ follows an Ornstein--Uhlenbeck process with mean reversion $\theta = 0.3$~h$^{-1}$ and volatility $\sigma = 0.1$, discretized hourly. The interannual spreads are deliberately conservative: satellite records give year-to-year coefficients of variation of 1--5\% for annual global horizontal irradiation over most of Kenya (5--6\% in the central and western highlands)~\cite{kariuki2018solar}, and wind resource studies report 3--6\% for annual mean wind speed, up to 10\% at some sites~\cite{lee2018wes}; the resulting scenario set has coefficients of variation of 5.1\% in annual irradiation, 9.9\% in mean wind speed and 0.3\% in annual load. All designs are evaluated on the same scenarios (common random numbers, seed 2026), with the diesel-only baseline fixed at its nominal value so that normalized objectives remain comparable across scenarios.

\begin{figure}[htbp]
  \centering
  \includegraphics[width=\linewidth]{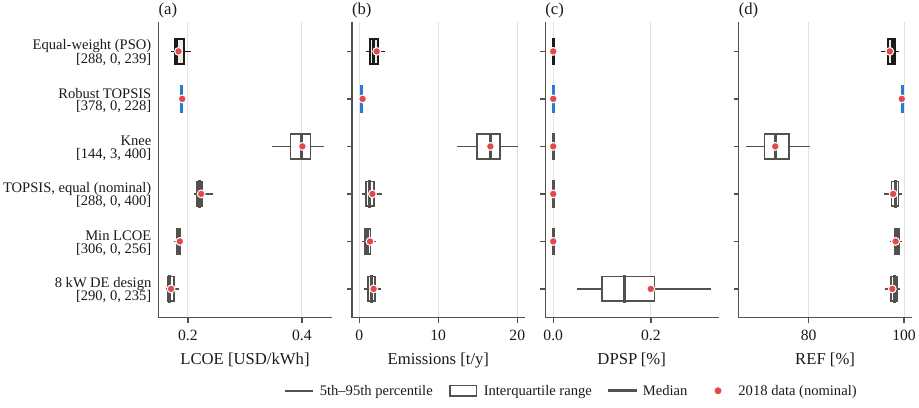}
  \caption{\label{fig:mc} Distributions of annual LCOE, emissions, DPSP and REF over 500 Monte Carlo scenarios for the designs of \cref{tab:mc}: lines span the 5th--95th percentiles, boxes the interquartile range, the bar marks the median and the dot the value for the 2018 data. The equal-weight design (black) and the robust TOPSIS choice (blue) are highlighted.}
\end{figure}

We evaluated the equal-weight design, the 8~kW DE design and all 200 points of the 16~kW Pareto approximation (\cref{sec:pareto_select}). \cref{tab:mc} and \cref{fig:mc} summarize six of them. For the equal-weight design, annual LCOE ranges from 0.172 to 0.205~USD/kWh (5th--95th percentile) with CVaR$_{95}$ = 0.213~USD/kWh, and the renewable fraction stays above 95.4\% in 95\% of scenarios. With the 16~kW DE, unmet load exceeds 1\% in none of the scenarios and is non-zero in 1.2\% of them, when load noise lifts an hourly peak above the DE rating while the battery is empty. With the 8~kW DE, some load is lost in every scenario, but DPSP stays below 0.3\% at the 95th percentile and never exceeds 1\%, at a lower LCOE (0.162--0.183~USD/kWh).

We then re-ranked the 16~kW candidates with TOPSIS (equal weights) on four risk-aware criteria: mean LCOE, CVaR$_{95}$ of LCOE, mean emissions and the probability of any unmet load. (The probability of exceeding 1\% unmet load is zero for every 16~kW candidate and therefore cannot discriminate between them.) The robust choice is [378, 0, 228], which differs from the equal-weight design: its mean LCOE is 2\% higher (0.189 vs 0.184~USD/kWh), but its CVaR$_{95}$ is 10\% lower (0.192 vs 0.213~USD/kWh), its mean emissions are 84\% lower and its probability of any unmet load falls from 1.2\% to 0.2\%. The 31\% larger PV array (96 vs 73~kW) reduces reliance on diesel in years with poor renewable resources, limiting high-cost outcomes. The equal-weight design ranks 87th out of 201 under this robust criterion. The knee point of the nominal front has the highest cost and emissions of the six, because it achieves a near-zero dump fraction with a smaller PV array that leaves more of the load to the DE. A designer who weighs cost tail risk and emissions above expected cost should therefore choose the robust design; one who prioritizes expected cost can keep the equal-weight design, whose LCOE distribution is narrow.

\begin{table}[htbp]
\centering
\scriptsize
\setlength{\tabcolsep}{3pt}
\begin{tabular}{lccccccc}
\toprule
Design & $[n_s, n_w, E_b]$ & LCOE nominal & LCOE 5--95\% & CVaR$_{95}$ LCOE & Emissions 5--95\% & $P$(DPSP$>$1\%) & REF mean\\
 & & [USD/kWh] & [USD/kWh] & [USD/kWh] & [kg/y] & & \\
\midrule
Weighted-sum PSO (chosen) & [288, 0, 239] & 0.184 & 0.172--0.205 & 0.213 & 920--3,192 & 0.00 & 0.973\\
Robust TOPSIS & [378, 0, 228] & 0.190 & 0.186--0.192 & 0.192 & 88--602 & 0.00 & 0.997\\
Knee & [144, 3, 400] & 0.401 & 0.350--0.438 & 0.447 & 12,490--19,945 & 0.00 & 0.732\\
TOPSIS (equal, nominal) & [288, 0, 400] & 0.223 & 0.212--0.243 & 0.253 & 430--2,743 & 0.00 & 0.980\\
Min LCOE & [306, 0, 256] & 0.186 & 0.177--0.192 & 0.202 & 452--2,027 & 0.00 & 0.985\\
8 kW DE design & [290, 0, 235] & 0.170 & 0.162--0.183 & 0.189 & 750--2,606 & 0.00 & 0.978\\
\bottomrule
\end{tabular}

\caption{\label{tab:mc} Monte Carlo performance of selected designs over 500 annual scenarios of irradiance, wind and load. LCOE nominal: value for the 2018 data; CVaR$_{95}$: mean LCOE over the worst 5\% of scenarios.}
\end{table}

\subsection{Deterministic stress tests\label{sec:stress}}
\begin{figure}[htbp]
  \centering
  \includegraphics[width=\linewidth]{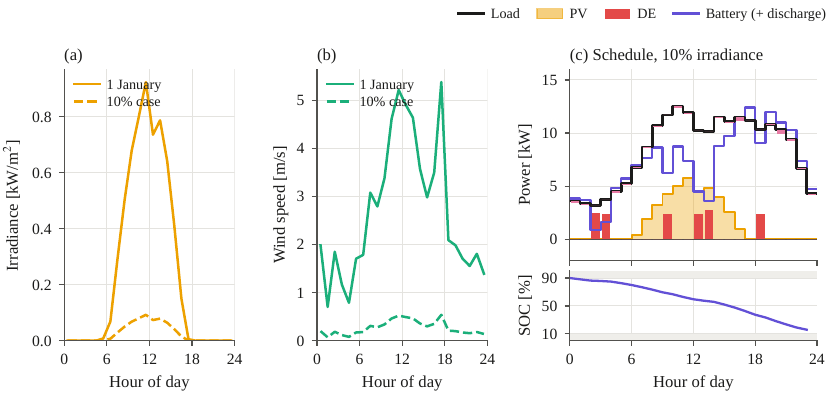}
  \caption{\label{fig:dispatch_robust_RES} Stress test with low renewable resource (8~kW DE design): (a) irradiance and (b) wind speed (measurement height) on the representative day and scaled to 10\%; (c) optimized schedule and SOC with 10\% irradiance. With 10\% wind speed alone the schedule is that of the representative day, and with both resources at 10\% it is that of (c), because the design has no wind turbines.}
\end{figure}

The Monte Carlo analysis describes typical variability; it does not test extreme single days. We therefore also test dispatch under deterministic resource and demand stresses, using the 8~kW DE design. The low-resource days scale the irradiance and wind speed of the representative day to 10\% of their values (\cref{fig:dispatch_robust_RES}a--b). They are deliberately extreme: the lowest-irradiation day of 2018 still delivers 31\% of the annual mean daily irradiation, and the calmest day, 17 June (\cref{sec:dispatch_opt}), has 43\% of the mean daily wind speed, so the 10\% cases represent severe stress conditions rather than typical days. \cref{tab:dispatch_robust_RES} shows that on a day with 10\% of the representative-day irradiance the DE runs for 6~h, the renewable fraction falls to 74\% and the daily COE rises to 0.133~USD/kWh, while DPSP stays within the 1\% limit. Low wind has no effect, because the design installs no wind turbines.

\begin{table}[htbp]
\centering
\small
\begin{tabular}{l c c c c c c c c}
\toprule
\textbf{Case} & \textbf{$J$} & \textbf{COE} & \textbf{COE} & \textbf{Em.} & \textbf{DPSP} & \textbf{Dump} & \textbf{REF} & \textbf{DE} \\
 & & [USD/kWh] & (norm.) & [kg] & [\%] & & [\%] & [h] \\
\midrule
Low solar irradiance (10\%) & 0.1728 & 0.133 & 0.3557 & 10.1 & 0.97 & 0.000 & 73.8 & 6 \\
Low wind speed (10\%) & 0.1686 & 0.057 & 0.1524 & 0.0 & 1.00 & 0.522 & 100.0 & 0 \\
Low solar and wind & 0.1728 & 0.133 & 0.3557 & 10.1 & 0.97 & 0.000 & 73.8 & 6 \\
\bottomrule
\end{tabular}
\caption{\label{tab:dispatch_robust_RES} Day-ahead dispatch under low renewable resource (8~kW DE design).}
\end{table}

\subsection{Demand profiles}
\begin{figure}[htbp]
  \centering
  \includegraphics[width=\linewidth]{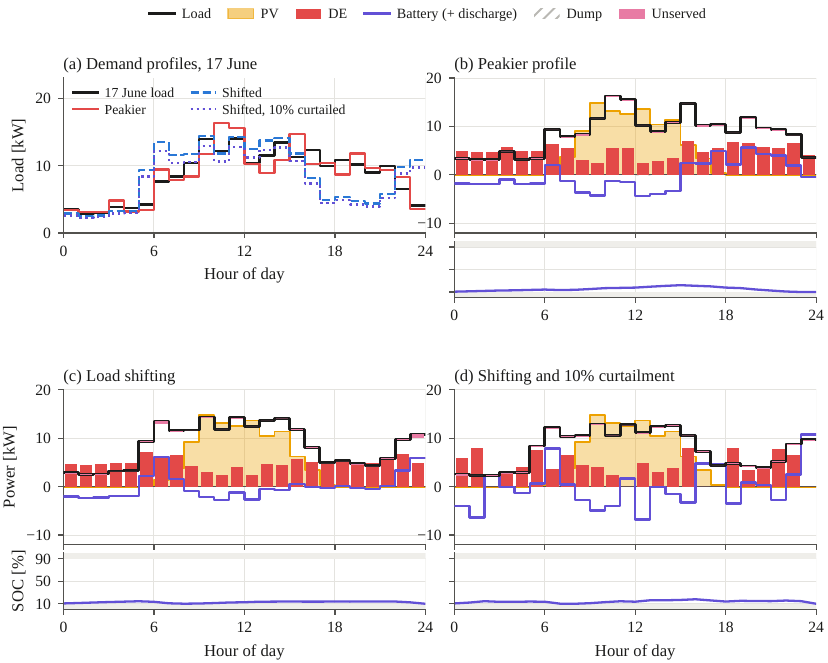}
  \caption{\label{fig:dispatch_load} Day-ahead dispatch of the 8~kW DE design on 17 June under alternative demand profiles: (a) the day's load and the three profiles derived from it; (b--d) optimized schedules, drawn as in \cref{fig:dispatch_P_base}.}
\end{figure}

We also repeated the dispatch with three alternative load profiles: a peakier profile in which hourly loads vary by up to $\pm 30\%$ around the baseline, a flatter profile in which load is shifted toward the hours of solar output, and the shifted profile with 10\% daily curtailment. On the sunny representative day, demand shape matters mainly through dumped energy: shifting load into daylight hours lowers the dump fraction from 0.522 to 0.458 and the weighted objective from 0.1686 to 0.1503, and the DE stays off in all cases (\cref{tab:dispatch_robust_load}). On 17 June we applied the same hour-by-hour changes to that day's load, preserving its total daily energy (90\% of that total with curtailment; \cref{fig:dispatch_load}a). With little solar surplus to absorb, shifting lowers the daily COE by only 1\% (0.241 vs 0.243~USD/kWh) and emissions by 1\%, while the peakier profile raises COE by 2\% and fuel use by 9\% (45.1 vs 41.5~L). Curtailing 10\% of the load cuts emissions by 17\% (66.7 vs 80.1~kg) and fuel use by 12\% (\cref{fig:dispatch_load}b--d). Demand-side measures that move load into hours of renewable surplus therefore provide the largest benefit on sunny days, while on low-resource days only a reduction in demand saves much diesel (\cref{sec:dr}).

\begin{table}[htbp]
\centering
\small
\begin{tabular}{l c c c c c c c c}
\toprule
\textbf{Case} & \textbf{$J$} & \textbf{COE} & \textbf{COE} & \textbf{Em.} & \textbf{DPSP} & \textbf{Dump} & \textbf{REF} & \textbf{DE} \\
 & & [USD/kWh] & (norm.) & [kg] & [\%] & & [\%] & [h] \\
\midrule
\multicolumn{9}{l}{\textit{1 January}} \\
Baseline profile & 0.1686 & 0.057 & 0.1524 & 0.0 & 1.00 & 0.522 & 100.0 & 0 \\
Peakier profile & 0.1682 & 0.057 & 0.1529 & 0.0 & 1.00 & 0.520 & 100.0 & 0 \\
Load shifting & 0.1503 & 0.053 & 0.1433 & 0.0 & 1.00 & 0.458 & 100.0 & 0 \\
Load shifting and 10\% curtailment & 0.1630 & 0.056 & 0.1435 & 0.0 & 1.00 & 0.509 & 100.0 & 0 \\
\midrule
\multicolumn{9}{l}{\textit{17 June}} \\
Baseline profile & 0.4357 & 0.243 & 0.6562 & 80.1 & 1.00 & 0.000 & 48.5 & 19 \\
Peakier profile & 0.4414 & 0.247 & 0.6673 & 81.2 & 0.97 & 0.001 & 48.2 & 24 \\
Load shifting & 0.4322 & 0.241 & 0.6507 & 79.3 & 1.00 & 0.000 & 48.8 & 24 \\
Load shifting and 10\% curtailment & 0.4066 & 0.244 & 0.6288 & 66.7 & 1.00 & 0.000 & 53.1 & 19 \\
\bottomrule
\end{tabular}
\caption{\label{tab:dispatch_robust_load} Day-ahead dispatch under alternative demand profiles (8~kW DE design) on the representative day and on the low-resource day.}
\end{table}

\section{Discussion and limitations\label{sec:discussion}}

\subsection{Site-specific and transferable findings}
Some findings are specific to the Timbila, Kenya case study and reflect the region's conditions. The PV--WT balance of the optimal mix follows from the local resource: 2018 irradiation at the site averaged 4.4~kWh/m$^2$/day in the station record (5.3~kWh/m$^2$/day in NASA POWER), while the wind resource is moderate and uncertain (\cref{app:climate}). In windier locations, such as coastal East Africa or the Sahel, or where the wind blows more at night, the balance would shift toward wind; \cref{sec:wind_role} gives the wind resource and turbine cost at which turbines enter the mix. The strong sensitivity of cost to diesel price reflects the elevated and volatile fuel costs of remote sub-Saharan sites; where fuel is cheaper or more stable, the economic case for high renewable penetration weakens.

Other findings depend on the structure of the problem rather than on the site. The design workflow (equal-budget multi-run benchmarking, Pareto search, rule-based selection from the front and Monte Carlo re-ranking) applies to any simulation-based sizing problem once the weather, load and cost inputs are replaced. The solver ranking in \cref{sec:results} is an empirical result for this problem class at a budget of 2,000 evaluations; other formulations or budgets may rank solvers differently, which is why we report the full protocol. The trade-off between generator rating and renewable and storage capacity, and the cost of tighter reliability targets traced by the $\varepsilon$-constraint front, hold for any islanded system that relies on a dispatchable backup unit. Practitioners at other off-grid sites can reuse the workflow and these relationships after recalibrating irradiance, wind, load, fuel price and financing inputs.

\subsection{Limitations\label{sec:limitations}}
The conclusions are subject to five modeling limitations. First, Stage~2 schedules the generator and battery with exact knowledge of the next 24 hours of load and weather. The dispatch costs, emissions and DPSP in \cref{sec:dispatch_opt} are therefore optimistic relative to a controller that acts on forecasts. Stage~1 does not share this bias, because its rule-based controller uses only current-hour information. Second, sizing and dispatch are decoupled (\cref{sec:opt_prob}): a design ranked best under rule-based operation may rank differently under optimized dispatch, although its Stage-1 cost bounds its cost under any better operating policy; for the selected design, on the day with the most diesel use, optimized dispatch is 0.4\% cheaper (\cref{sec:dispatch_opt}). Third, sizing uses one meteorological year (2018) and one load profile. The Monte Carlo analysis perturbs both, but it cannot represent multi-year wind droughts or long-term load growth. Fourth, the wind resource rests on station data scaled by a single factor (\cref{app:climate}); \cref{sec:sizing_sensitivity} shows how the design responds when that factor changes by $\pm$25\%, and on-site hub-height measurements would be needed before construction. Fifth, solvers use default settings except for the changes listed in \cref{tab:protocol}. Further tuning could change their relative performance. The common evaluation budget caps computational effort, while individual stopping rules may end runs earlier.

\section{Conclusions}

We sized an islanded PV, wind, Li-ion and diesel microgrid for an off-grid community in Timbila, Kenya, using a workflow that combines a solver benchmark with statistical tests, a posteriori Pareto search with explicit selection rules, and Monte Carlo re-ranking of candidate designs. The main findings are as follows.

First, on this non-smooth, simulation-based problem, PSO, the PSO--pattern-search hybrid, pattern search and surrogate optimization finished within 0.5\% of the exhaustively computed integer optimum in all 30 runs at a budget of 2,000 evaluations, and PSO did so with the lowest expected running time (2.9~s); the gradient-based multistart methods were the least reliable.

Second, the equal-weight design installs 73~kW of PV, a 239~kWh Li-ion battery and no wind turbines. It supplies 97\% of the energy from renewables at 0.184~USD/kWh, 63\% below a diesel-only system, with emissions 96\% lower. The generator remains necessary for reliability: with the generator rating as a decision variable, equal weights select no generator and accept 11\% unmet load, whereas the $\varepsilon$-constraint front shows that a design with a 12.2~kW generator serves all load at 0.183~USD/kWh. The reliability target sets the size of the generator, while the PV array and battery hardly change. Wind turbines do not enter the design because the site's wind blows mostly in daylight, together with the PV output; including four turbines raises LCOE by 4.6\%, and turbines would become economical only with a substantially stronger wind resource or cheaper small turbines.

Third, selection rules and uncertainty change which design a planner should build. Knee-point, compromise and TOPSIS rules pick designs that differ in battery size and cost, and the weighted-sum design is not dominated by any point of the Pareto approximation. Over 500 Monte Carlo scenarios the equal-weight design never loses more than 1\% of load, but a design with a 31\% larger PV array lowers the 95\% conditional value-at-risk of LCOE by 10\% at a 2\% higher expected cost, and cuts mean emissions by 84\%.

Fourth, the economics depend strongly on financing. LCOE ranges from 0.097 to 0.366~USD/kWh as the nominal interest rate varies from 0 to 20\%, far more than across the diesel price range (0.161--0.241~USD/kWh), which makes concessional finance important for the affordability of off-grid renewables in sub-Saharan Africa. On the day of 2018 with the most diesel use, optimized day-ahead dispatch lowers the selected design's cost by less than 1\% relative to the rule-based controller used for sizing; with a smaller 8~kW generator, optimized dispatch keeps unmet load within 1\%. Shifting load into daylight hours reduces dumped energy on sunny days, while on low-resource days only curtailment saves much diesel.

These results rest on one year of measured weather, a single wind scaling factor and day-ahead dispatch with perfect foresight; \cref{sec:limitations} discusses these limits and \cref{sec:future} the extensions that would address them.

\subsection{Future work \label{sec:future}}

The limitations in \cref{sec:limitations} suggest three extensions: integrated or bilevel sizing and dispatch, in which designs are evaluated under optimized operation; receding-horizon or stochastic dispatch based on forecasts; and multi-year resource records with on-site wind measurements. Adaptive and predictive control, improved load and renewable forecasts, and demand-response programs could improve performance under uncertain conditions (\cref{app:improv}).

The model currently considers electricity only. It could be extended to combined heat and power (CHP) or combined cooling, heating and power (CCHP) microgrids with both thermal and electrical loads. Other storage technologies include electrical, mechanical, thermal and magnetic storage \cite{ESS_control}. Small-scale pumped hydro and micro-hydropower schemes are also options for rural Africa. Alternatives to fossil-fueled backup generation include hydrogen fuel cells and biomass-fueled generation; biomass is abundant in many African countries \cite{usea}.

\section{Data and code availability}

All code and data used for this study are in the public GitHub repository \url{https://github.com/vineetjnair9/microgrid_optimization}. Version \texttt{v2.0-asoc} of the \texttt{master} branch (folder \texttt{revision\_ASOC}) contains the simulation model, the benchmark, Pareto, sensitivity, Monte Carlo and dispatch scripts, the random seeds, all result files used in this paper and the scripts used to generate the figures.

\section{Acknowledgements}
The author thanks Prof. Ioannis Lestas at the University of Cambridge for helpful discussions and guidance on the project. The author was supported by a Gates Cambridge Scholarship. There are no conflicts of interest to report.

\newpage

\appendix
\crefalias{section}{appendix}
\crefalias{subsection}{appendix}
\crefalias{subsubsection}{appendix}

\section{Microgrid system models \label{app:mg_models}}

\subsection{Renewable energy sources (RES)}

Renewable power in the region relies heavily on biomass and large hydropower, which we exclude. Biomass such as firewood and charcoal is carbon-neutral but harms vegetation and biodiversity, is land-intensive, drives deforestation and competes with food crops. Large hydroelectric schemes can damage local ecosystems and displace communities. Most suitable sites in East African countries such as Kenya and Tanzania are already developed, and recurring droughts reduce reservoir head and flow, limiting hydropower output \cite{usea}. Wind turbines and solar PV are therefore the only renewables considered.

\subsubsection{Solar PV arrays}

The hourly DC power output [$kW$] of the PV generator is calculated as $P_{PV} = n_s \; \eta_{PV} \; A_c \; I_{PV}$ \cite{tazvinga2013}, where $n_s$ is the number of PV modules in the system, $\eta_{PV}$ is the power conversion efficiency of each module, expressed as a fraction, $A_c$ is the collector area of each module [$m^2$], and $I_{PV}$ [$kW/m^2$] is the hourly mean global horizontal irradiance (GHI) used as the irradiance incident on the array. PV module efficiency varies as a function of incident radiation $I_{PV}$, ambient air temperature $T_a$, and cell parameters measured under nominal operating cell temperature (NOCT) and standard test conditions \cite{tazvinga2013}:

\begin{equation}
\label{eq:pv2}
    \eta_{PV} = \eta_r \eta_{pc} \cdot \left(1 - 0.9 \beta \left(\frac{I_{PV}}{I_{PV,NOCT}}\right)(T_{c,NOCT} - T_{a,NOCT}) -\beta(T_a - T_r)\right)
\end{equation}

\noindent where $\eta_r$ is the PV efficiency measured at reference cell temperature $T_r$ and standard test conditions. $\beta$ is the temperature coefficient of cell efficiency which tracks degradation in cell performance with rising temperature. $I_{PV,\;NOCT}$ is the irradiance under NOCT test conditions, while $T_{c,\;NOCT}$ and $T_{a,\;NOCT}$ are the cell and ambient temperatures at NOCT test conditions, respectively. The power conditioner performs maximum power point tracking (MPPT); assuming perfect MPPT, the power conditioning efficiency is $\eta_{pc}= 1$ \cite{kaabeche2011b}. We use Mitsubishi Electric monocrystalline silicon modules rated at 255~W (\cref{tab:pv_cell}).

\begin{table}[htbp]
\centering
\begin{tabular}{@{}cccccccc@{}}
\toprule
$\eta_r$ [\%]  & $T_r$ [$\degree$ C] & $I_{PV,NOCT}$ [$kW/m^2$]  & $T_{c,NOCT}$ [$\degree$ C]  & $T_{a,NOCT}$ [$\degree$ C] & $\beta$ [\%/$\degree$ C] & $P_{PV,\;r}$ [$kW$] & $A_c$ [$m^2$] \\ \midrule
15.4 & 25 & 0.8 & 45.7 & 20 & 0.0045 & 0.255 & 1.4602
\\ \bottomrule
\end{tabular}
\caption{Specifications of the 255 Wp PV-MLU255HC model \cite{mitsubishi}.}
\label{tab:pv_cell} 
\end{table}

\subsubsection{Wind turbines (WT)}

We use small onshore horizontal-axis turbines from Raum Energy rated at 3.5~kW each (five blades, downwind). A small rating suits the low demand of the rural community and keeps installation and permitting fees, and hence the upfront capital cost, low. \cref{tab:wt_spec} lists the specifications. Turbine output depends on the wind speed at the site, the air density $\rho_{air}$, the swept area $A_{swept}$ and the efficiency $\eta_w$ with which the rotor converts the kinetic energy of the wind into electricity \cite{tazvinga2014}:

\begin{equation}
\label{eq:Pwind}
    P_{WT} = n_w \; \frac{1}{2} \; \eta_w \; \rho_{air} \; C_p \; A_{swept} \; v_{hub}^3, \quad v_{hub}(t) = v_{ref}(t) \; \left(\frac{h_{hub}}{h_{ref}}\right)^{\alpha_w}
\end{equation}

\noindent where $n_w$ is the number of turbines and $v_{hub}$ is the wind speed at hub height $h_{hub}$, extrapolated from the hourly wind speed $v_{ref}$ measured at the reference height $h_{ref}$ close to the ground. The power-law exponent $\alpha_w$ ranges from below 0.1 over flat land, water or ice to above 0.25 over forest \cite{kaabeche2011a}; we use 0.14, which suits open grassland away from tall trees and buildings, as around rural villages in Kenya and Tanzania \cite{kaviani2009}. A variable-speed permanent-magnet generator keeps the rotor at the tip-speed ratio $\lambda$ that maximizes the power coefficient $C_p$ and the conversion efficiency $\eta_w$, and hence the power output \cite{seeling1997}.

\begin{table}[htbp]
\centering
\begin{tabular}{@{}cccccccc@{}}
\toprule
\textbf{Parameter} & $h_{hub} \; [m]$ & $h_{ref} \; [m]$ & $P_{WT,r} \; [kW]$ & $A_r \; [m^2]$ & $v_c \; [m/s]$ & $v_r \; [m/s]$ & $v_f \; [m/s]$ \\ \midrule
\textbf{Value}    & 14.5              & 1                  & 3.5                    & 12.6                                                & 2.8                & 11                                         & 22 \\ \bottomrule
\end{tabular}
\caption{Specifications of the small-scale WT system \cite{raum}.}
\label{tab:wt_spec}
\end{table}

The manufacturer's data sheet gives neither $\eta_w$ nor $C_p$, nor their variation with wind speed, so we compute the output with an empirical power curve with three regions, based on the total rated power $P_{WT,\;r,\; total} = n_w \cdot P_{WT,\;r}$ and $v_{hub}$:

\[
 P_{WT}(v_{hub}) = 
  \begin{cases} 
   n_w \cdot P_{WT,\;r} \cdot (A + B\cdot v_{hub} + C \cdot v_{hub}^2) & \text{if } v_c \leq v_{hub} \leq v_{r} \\
   n_w \cdot P_{WT,\;r}      & \text{if } v_{r} \leq v_{hub} \leq v_{f} \\
   0 & \text{otherwise}
  \end{cases}
\]
where $v_c$, $v_r$, and $v_f$ are the cut-in (start-up), rated, and cut-out (shutdown) wind speeds, respectively. The constants $A$, $B$ and $C$ can be computed using \cite{wen2009}:

\begin{align*}
    A & = \frac{1}{(v_c - v_r)^2} \left(v_c(v_c + v_r) - 4 v_c v_r \left(\frac{(v_c + v_r)}{2v_r}\right)^3 \right), \quad B = \frac{1}{(v_c - v_r)^2} \left(4 (v_c + v_r) \left(\frac{(v_c + v_r)}{2v_r}\right)^3 - (3v_c + v_r)\right) \\
    C & = \frac{1}{(v_c - v_r)^2} \left(2 - 4 \left(\frac{(v_c + v_r)}{2v_r}\right)^3 \right)
\end{align*}

\subsubsection{Climate data \label{app:climate}}

\begin{figure}[htbp]
  \centering
  \includegraphics[width=\linewidth]{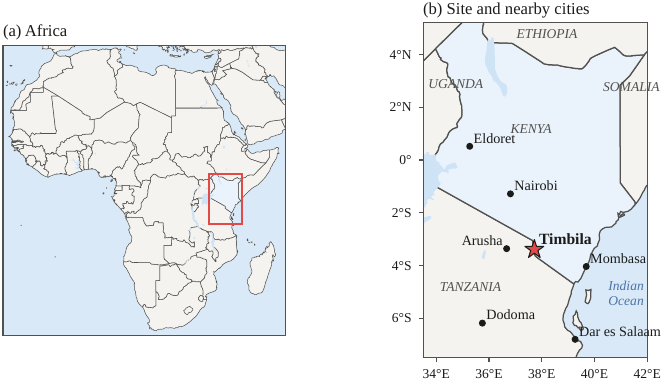}
  \caption{Location of the microgrid site near Timbila, Kenya: (a) the region within Africa; (b) the site and nearby cities. Boundaries, lakes and places from Natural Earth (public domain, naturalearthdata.com).}
  \label{fig:chap3_timbila}
\end{figure}

Solar and wind output are computed from hourly temperature, wind speed and solar irradiance for 2018 measured by ground stations of the Trans-African HydroMeteorological Observatory (TAHMO), a public-benefit network of weather stations across Africa \cite{tahmo}; ground measurements tend to be more accurate than satellite estimates, on average. The microgrid serves a rural village near Timbila in southern Kenya, close to the Tanzanian border (\cref{fig:chap3_timbila}), a site chosen because weather and load data were available. Missing wind speed values at the Timbila station were filled with the average of the four neighboring TAHMO stations listed in \cref{tab:tahmo_stations}.

\begin{table}[htbp]
\centering
\begin{tabular}{p{9cm} p{1cm} p{1cm} p{1cm} p{1cm}}
\toprule
\multicolumn{1}{l}{\textbf{Location}}    & \multicolumn{1}{c}{\textbf{Lat (\degree N)}} & \multicolumn{1}{c}{\textbf{Lon (\degree E)}} & \multicolumn{1}{c}{\textbf{Elevation (m)}} \\ \midrule
Timbila High School, Kenya                                        & -3.39                                     & 37.72                                      & 743                                        \\
Imurtott Primary School, Kenya                                      & -2.88                                      & 37.39                                       & 1853                                       \\
South Eastern Kenya University - Mtitu Adei Campus             & -2.66                                     & 38.11                                      & 854                                        \\
Our Lady of Perpetuah Girls Secondary School, Kenya                   & -3.50                                       & 38.59                                      & 1158                                       \\
Bangalala, Tanzania                                                 & -4.23                                     & 37.85                                      & 877                                        \\ \bottomrule
\end{tabular}
\caption{Geographical details of stations used to generate combined climate data \cite{tahmo}.}
\label{tab:tahmo_stations}
\end{table}

\begin{figure}[htbp]
  \centering
  \includegraphics[width=\linewidth]{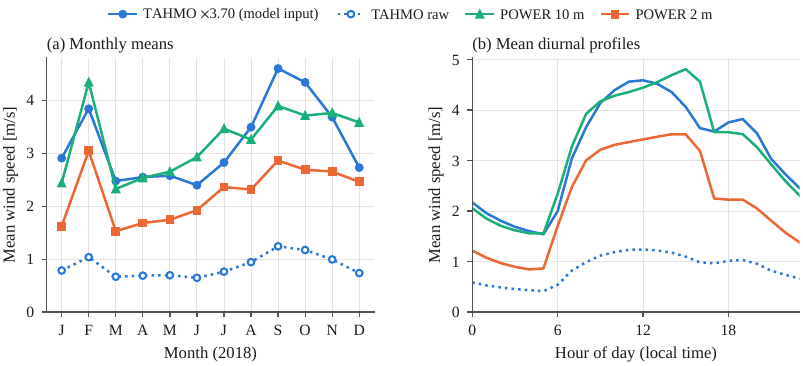}
  \caption{\label{fig:wind_val} TAHMO station wind (raw and scaled by 3.70) compared with NASA POWER (MERRA-2) wind at 2~m and 10~m for Timbila, 2018: (a) monthly means, (b) mean diurnal profiles after time alignment.}
\end{figure}

The TAHMO wind speeds at the Timbila station average only 0.86~m/s, far below typical values for the area. We therefore scaled the series by a factor of about 3.70 (= 3.19/0.86) so that its annual mean matched forecast-website values for Timbila. We checked the series against an independent source, NASA POWER hourly data for 2018 at the same coordinates \cite{nasapower}, whose wind speeds come from the MERRA-2 reanalysis \cite{gelaro2017merra2}. The station time stamps lead local time: the irradiance series match best after a 2-hour shift ($r = 0.96$), and the load profile is in local time, so we shift all station series (irradiance, temperature and wind) by 2~h. After this alignment the two wind series agree in timing ($r = 0.63$ hourly, 0.60 for daily and 0.82 for monthly means; \cref{fig:wind_val}) but not in level: the reanalysis 2~m wind averages 2.23~m/s, 2.59 times the TAHMO mean, and monthly ratios range from 2.07 to 3.33 (\cref{tab:wind_val}). At the 14.5~m hub height, the scaled TAHMO data with the model's 1~m reference height give a mean of 4.68~m/s, while the reanalysis 10~m wind extrapolated with its own 10--50~m shear gives 3.50~m/s, 25\% lower. TAHMO stations use the METER ATMOS~41 all-in-one sensor~\cite{meter_tahmo}, which is commonly mounted 2~m above ground~\cite{tahmo_install}; the mounting height of the Timbila sensor is not recorded in the data we used. A 2~m reference height would lower the model's hub-height mean by 9\% (4.24~m/s). Reanalysis data represent a grid cell of about 50~km in complex terrain near Mount Kilimanjaro rather than the site itself, so neither source is definitive; on-site measurement at hub height would be needed before construction. We retain the 3.70 scaling factor and report the sensitivity of the design to the wind level in \cref{sec:sizing_sensitivity}: the optimal design installs no wind turbines whether the wind input is scaled by 0.75 (which matches the reanalysis hub-height mean), 1.00 or 1.25, so the uncertainty in the wind data does not change the result. The TAHMO irradiance agrees closely with the CERES irradiance in NASA POWER in timing ($r = 0.96$) and is 16\% lower on average (4.42 vs 5.28~kWh/m$^2$/day), so the station data provide a lower estimate of solar availability.

\begin{table}[htbp]
\centering
\scriptsize
\setlength{\tabcolsep}{3pt}
\begin{tabular}{lrrrrrrrrrrrr}
\toprule
Month & J & F & M & A & M & J & J & A & S & O & N & D\\
\midrule
TAHMO raw [m/s] & 0.79 & 1.04 & 0.67 & 0.69 & 0.70 & 0.65 & 0.76 & 0.94 & 1.24 & 1.17 & 0.99 & 0.74\\
POWER 2 m [m/s] & 1.62 & 3.04 & 1.53 & 1.68 & 1.74 & 1.92 & 2.36 & 2.31 & 2.86 & 2.69 & 2.66 & 2.46\\
Ratio & 2.07 & 2.93 & 2.28 & 2.44 & 2.50 & 2.97 & 3.09 & 2.45 & 2.30 & 2.29 & 2.67 & 3.33\\
\bottomrule
\end{tabular}

\caption{\label{tab:wind_val} Monthly mean wind speed of the raw TAHMO series and NASA POWER at 2~m, and their ratio (2018).}
\end{table}  

\begin{figure}[htbp]
  \centering
  \includegraphics[width=\linewidth]{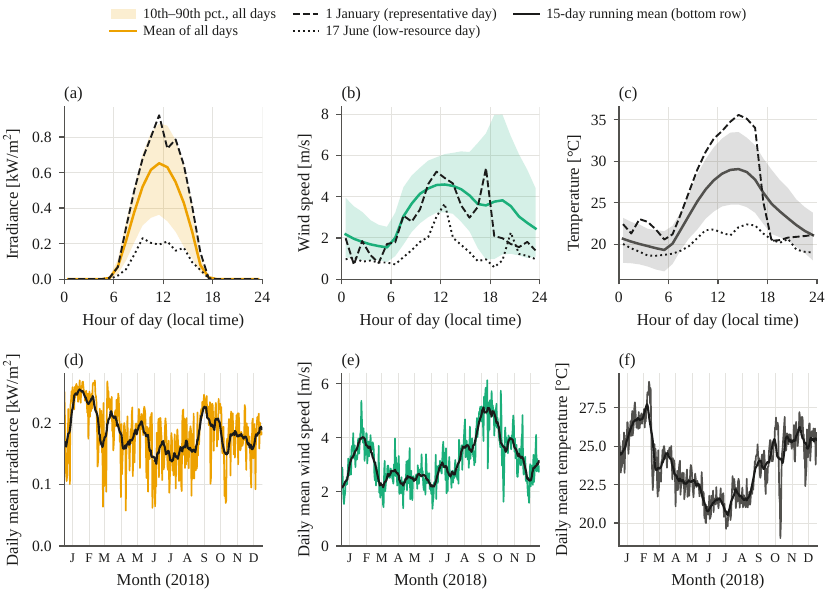}
  \caption{\label{fig:climate} Climate inputs for 2018 after alignment to local time. Top: mean diurnal profiles of (a) irradiance, (b) wind speed (station data scaled by 3.70, 1~m reference height) and (c) ambient temperature, with the 10th--90th percentile band over all days and the profiles of 1 January and 17 June, the two days used for dispatch in \cref{sec:dispatch_opt}; bottom (d--f): daily means with a 15-day running mean.}
\end{figure}

\cref{fig:climate} summarizes the aligned series used in the model.

\subsection{Electricity demand (load)}

\begin{figure}[htbp]
  \centering
  \includegraphics[width=\linewidth]{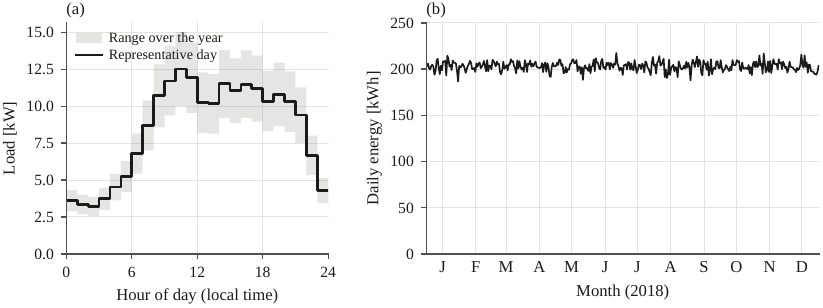}
  \caption{Load: (a) representative-day profile (1 January; household and commercial demand from REopt \cite{nrel_data}) and the range of hourly load over the year; (b) daily energy over 2018.}
  \label{fig:chap3_load}
\end{figure}

The load is a characteristic rural African profile from NREL's REopt tool \cite{nrel_data} for the community described in \cref{tab:load}. We neglect differences between weekdays and weekends and between seasons, which is reasonable for a largely residential village with few businesses in Timbila's tropical climate, where conditions are similar all year. The representative-day profile, the sum of household and commercial demand, peaks at 12.52~kW between 10:00 and 11:00 (\cref{fig:chap3_load}a). To build the annual series, we use this profile for 1 January 2018 and multiply each hourly value on the other 364 days by an independent factor drawn uniformly from $[0.8, 1.2]$. Hourly load therefore varies by up to $\pm20\%$ around the profile and peaks at 15.0~kW over the year, while daily energy stays between 187 and 217~kWh (\cref{fig:chap3_load}b). \cref{tab:load_summary} summarizes the series.

\begin{table}[htbp] 
\centering
\begin{tabular}{@{}lc@{}}
\toprule
\multicolumn{1}{c}{\textbf{Parameter}} & \textbf{Value}       \\ 
\midrule
Number of households & 100 \\
High-income households & 33\% \\
Middle-income households & 33\% \\
Low-income households & 33\% \\
Number of water-pumping operations & 6 \\
Number of milling operations & 4 \\
Number of small shops & 10 \\
Number of schools & 1 \\
Number of clinics & 3 \\
Number of street lights & 30 \\
\bottomrule
\end{tabular}
\caption{Description of the community served by the microgrid near Timbila, Kenya.}
\label{tab:load}
\end{table}

\begin{table}[htbp]
\centering
\begin{tabular}{@{}cccccc@{}}
\toprule
Mean daily energy & Annual energy & \multicolumn{3}{c}{Representative day [kW]} & Annual peak \\
\cmidrule(lr){3-5}
[kWh/day] & [kWh/y] & Peak & Minimum & Mean & [kW] \\ \midrule
203 & 74,086 & 12.52 & 3.21 & 8.48 & 15.0 \\ \bottomrule
\end{tabular}
\caption{Summary statistics of the load.}
\label{tab:load_summary} 
\end{table}

\subsection{Backup distributed generators (DG) \label{sec:DG}}

Standby fossil-fueled generators supply the demand that renewables and storage cannot meet, for example after sudden increases in demand or during poor weather. As for the wind turbines, we assume variable-speed operation, which uses less fuel at part load than constant-speed operation because the engine or turbine slows down at low demand \cite{seeling1997}. To keep the problem simple, the system has a single DG. In the sizing problem the rating of the diesel engine or microturbine is fixed; \cref{sec:pareto_rel} treats the DE rating as a decision variable. The rating covers the peak demand with a margin, $P_{DG,\;r} = P_{load,\;peak} \cdot (1 + SF)$ \cite{ghasemi2013}, where the safety factor $SF$ sets the degree of oversizing; the margin acts as spinning reserve.

\subsubsection{Diesel engines (DE)}

Diesel engines are reciprocating compression-ignition engines. Unlike the RES they are dispatchable, start quickly and follow load well, but they have high fuel costs and emit greenhouse gases, particulates and noise \cite{mg_review}. The load peaks at 12.52~kW on the representative day used for dispatch and at 15.0~kW over the year (8 June, 10:00--11:00). The DE is rated at 16~kW, 28\% above the representative-day peak ($SF \approx 0.278$) and 6.6\% above the annual peak. According to \cite{fathima2015}, the lower limit on the DE's power output can be estimated as $P_{DE,\;min} = 0.3 \cdot P_{DE,\; r} = 4.8 \; kW$. In every hour in which the DE is committed ($DE_{ON}(t) = 1$), its volumetric fuel consumption follows the linear law $Fuel_{DE}(t) = \alpha_{DE} P(t)_{DE,\;gen} + \beta_{DE} P(t)_{DE,\;r}$, where $Fuel_{DE}(t)$ is in $L/h$, $P(t)_{DE,\;gen}$ [$kW$] is the power output needed by the load and $P(t)_{DE,\;r}$ [$kW$] is the rated power \cite{kaabeche2014}, $\alpha_{DE}$ [$L/kWh$] and $\beta_{DE}$ [$L/kWh$] are fuel consumption curve coefficients supplied by the manufacturer, assumed to be 0.246 and 0.08145 respectively \cite{kaabeche2014}. The hourly fuel costs of the DE can be calculated assuming a unit price of \$ 3.20 per US liquid gallon of diesel, as $C_{fuel,\; DE}(t) = 3.20 \cdot \frac{Fuel_{DE}(t)}{3.78541}$. The DE consumes no fuel in hours when it is off.

\subsubsection{Micro-gas turbines (MT)}

A microturbine can run on a wide range of gaseous and liquid fuels and suits combined heat and power (CHP) applications, for which natural gas is the most common fuel. Microturbines emit less greenhouse gas than DEs, and much less than large central natural gas plants, especially $NO_x$ and $CO_2$ \cite{wu2014}, but they cost more upfront, often have higher fuel costs than diesel, and are far less common in Africa than diesel gensets. They usually range from 30 to 330~kW; for a direct comparison we use the same rating as the DE, $P_{MT,\; r} = P_{MT, \; max} = 16$~kW with $P_{MT,\;min} = 4.8$~kW. A 61~kW MT burns 0.84~MMBtu per hour at rated power \cite{chp_mt}; scaling this, the hourly fuel use [MMBtu/h] at output $P_{MT}$ is $Fuel_{MT}(t) = \frac{0.84}{61} \cdot P_{MT}(t)$, and at a natural gas price of $C_{NG} = 2.19$~USD/MMBtu \cite{ng_price} the hourly fuel cost is $C_{fuel,\;MT}(t) = 2.19 \cdot Fuel_{MT}(t)$.

\subsubsection{Converter}

The bidirectional converter operates as both an inverter and a rectifier. Its power rating equals the representative-day peak load (12.52~kW). The converter is costed at this rating, but its power limit is not enforced in the hourly simulation. Both AC--DC and DC--AC conversion efficiencies are assumed to be 90\%, i.e., $\eta_{inv} = \eta_{rec} = 90\%$ \cite{ogunjuyigbe2016}. The assumed lifetime is 20~y \cite{moshi2016}, and the capital cost is $IC_{conv} = \$2800$ \cite{conv_price}.

\subsection{Battery energy storage (BS) \label{sec:BS}}

We consider lithium-ion and lead-acid batteries, using the same discrete-time model with technology-specific parameters to describe charging and discharging \cite{tazvinga2014}:
\begin{equation}
    SOC(t+1) = (1-\delta) SOC(t) - \frac{1000 \cdot P_{BS} \; \Delta t \; \eta_{BS}(t)}{E_C}, \qquad \eta_{BS}(t) = \begin{cases} \eta_{overall} & P_{BS} < 0 \text{ (charging)} \\ 1 & P_{BS} \geq 0 \text{ (discharging)} \end{cases}
\end{equation}
where $P_{BS}$ [$kW$] is the DC power dispatched either to or from the battery in each time step $\Delta t$ and $\delta$ is the hourly self-discharge rate (\%/h). A positive $P_{BS}$ value indicates that the battery is being discharged while a negative $P_{BS}$ implies charging. $SOC$ is the state of charge, expressed as the fraction of the current maximum energy capacity $E_C$, i.e., $SOC = \frac{E_{releasable}}{E_C}$. Since it is difficult to measure both charging and discharging efficiencies separately, an overall round-trip Coulombic efficiency $\eta_{overall}$ for the battery is used and applied once, when charging. In each hour the controller limits charging to the headroom below $SOC_{max}$ and discharging to the energy above $SOC_{min}$, in addition to the power rating $P_{BS,\;r}$. The power rating scales with installed capacity, as for stacked modules: 3.68~kW per 13.5~kWh Powerwall module for Li-ion (0.273~kW per kWh) and a C-rate of 0.2 for lead-acid. The parameters for both battery types are listed in \cref{tab:bs_params}.

\begin{table}[htbp]
\centering
\begin{tabular}{@{}lll@{}}
\toprule
\textbf{Parameter}               & \textbf{Li-ion (LI)} & \textbf{Pb-acid (LA)} \\ \midrule
$SOC_{min}$                         & 10\%            & 50\%             \\
$SOC_{max}$                         & 90\%            & 90\%             \\
$\delta$ [\%/month] & 7.5             & 5                \\
$\eta_{overall}$     & 90\%            & 75\%               \\
Lifetime [in y]                  & 15              & 5                \\
Lifetime $t_{BS}$ [in cycles]             & 5475            & 1400             \\
$P_{BS, \; r}$ [kW per kWh installed]  & 0.273           & 0.20             \\
Capacity fading $\Delta_E$ [\%/cycle]       & 0.0055          & 0.0214           \\ \bottomrule
\end{tabular}
\caption{Parameters of the Li-ion and lead-acid BS systems \cite{bs_source}.}
\label{tab:bs_params} 
\end{table}

The SOC must stay between $SOC_{min}$ and $SOC_{max}$. The simulation starts with the battery at $SOC_{max} = 90\%$ for both LI and LA. The lower limit follows from the depth of discharge $DoD$, the deepest level to which the battery can be safely discharged: $E_{b,\;min} = (1 - DoD)\cdot E_C$. The maximum capacity $E_C$ starts at the rated capacity $E_{b,\; init}$ and fades linearly with cycling, $E_C = E_{b,\; init} \cdot (1 - n_{cycles}\cdot \Delta_E)$, where $n_{cycles}$ is the cumulative number of equivalent full cycles, computed as the energy discharged divided by the usable capacity $(SOC_{max}-SOC_{min})\,E_{b,\;init}$ and updated every time step; over the one-year simulation this gives the number of cycles per year. The fade rates in \cref{tab:bs_params} assume that the battery is replaced when $E_C$ falls to 70\% of its rated value.
Some Li-ion parameters follow the Tesla Powerwall \cite{powerwall}, assuming that the battery can be fully cycled daily for five years beyond its 10-year warranty \cite{bs_source}.

Lead-acid batteries have shorter lifetimes, requiring more frequent replacement, and lower round-trip efficiency than Li-ion (\cref{tab:bs_params}). Li-ion also allows deeper discharge, so more of its capacity is usable, and its higher power rating per kWh lets it serve more of the demand. It has slightly higher self-discharge, mainly from its protection circuit, and costs more per kWh, but its slower capacity fade means fewer replacements, which can make it cheaper over the project life, depending on how often it is cycled.

\section{Details of cost functions in the objective \label{app:costs}}

The project lasts $t_{sys} = 25$~y, so all cash flows are discounted to the present; \cref{tab:financial} lists the financial assumptions.

\begin{table}[htbp]
\centering
\begin{tabular}{@{}lcc@{}}
\toprule
\multicolumn{1}{c}{\textbf{Parameter}} & \textbf{Symbol}      & \textbf{Value}       \\ \midrule
Nominal interest or discount rate (\%) & $i$                    & 9 \cite{interest_rate}                   \\
Inflation or escalation rate (\%)                    & $f$                    & 5.7 \cite{inflation_rate}                 \\
System lifetime (y)                    & $t_{sys}$           & 25   \\
PV lifetime & $t_s$ & 25 \\
WT lifetime & $t_w$ & 20 \\
\bottomrule
\end{tabular}
\caption{Key financial parameters used in the model.}
\label{tab:financial} 
\end{table}

\subsection{Sizing optimization problem costs}

Initial capital costs (IC) of the PV, WT and DGs scale with rated power and those of the batteries with rated energy capacity. Replacement costs (RC) are a fraction of IC because they exclude one-time permitting and installation fees. PV, WT and batteries have only fixed annual operation and maintenance (O\&M) costs, set as a fraction of IC, and the converter has none. The DGs also incur fuel costs, variable O\&M per operating hour, and startup ($SUC$) and shutdown ($SDC$) costs. \cref{tab:costs1,tab:costs2} summarize the values.

\begin{table}[htbp]
\centering
\begin{tabular}{@{}lcccc@{}}
\toprule
\textbf{Parameter}        & PV   & WT   & Li-ion & Pb-acid \\ \midrule
$IC_{kW}$ [USD/kW]           & 1210 & 1500 & -      & -       \\
$IC_{kWh}$ [USD/kWh]           & -    & -    & 300    & 255     \\
$RC$ [\% of $IC$]         & -    & -    & 90     & 90      \\
$OM_{fix}$ [\% of $IC$/y] & 1    & 3    & 1      & 1       \\ \bottomrule
\end{tabular}
\caption{Costs assumed for the RES and BS.}
\label{tab:costs1}
\end{table}

\begin{table}[htbp]
\centering
\begin{tabular}{@{}lcc@{}}
\toprule
\textbf{Parameter} & DE \cite{moshi2016}    & MT \cite{chp_mt}    \\ \midrule
$IC_{kW}$ [USD/kW]     & 781.25 & 3320  \\
$RC$                 & 88     & 90    \\
$OM_{fix}$ [\% of $IC$/y]          & 2      & 2     \\
$OM_{var}$            & 0.24 [USD/h]   & 0.013 [USD/kWh] \\
$SUC$ [\$]                & 0.45   & 0.45  \\
$SDC$ [\$]               & 0.23   & 0.23  \\
Lifetime $t_{DG}$ [h]  & 15000  & 40000 \\ 
\bottomrule
\end{tabular}
\caption{Cost assumptions for the DGs.}
\label{tab:costs2}
\end{table}

The WT and PV capital costs are global weighted averages of installed costs \cite{costs_irena}. Fixed O\&M costs are taken from \cite{kaabeche2011a}, and battery costs from \cite{hu2017}. The WT and PV capital costs may underestimate those of small microgrids because the averages include utility-scale projects that benefit from economies of scale. Replacement costs include only the batteries and backup DGs. The converter and WT are excluded because their lifetimes are close to the 25-year project lifetime.

The binary vectors $\vv{\bm{DG_{ON}}}(t)$ (i.e., $\vv{\bm{DE_{ON}}}(t)$ and $\vv{\bm{MT_{ON}}}(t)$) record the DG's operating state in each interval $t$ for calculating startup, shutdown and variable O\&M costs. $DG_{ON}(t) = 0$ denotes an offline generator and $DG_{ON}(t) = 1$ an online generator. The binary variable $BS_{cycled}(t)$ records whether the battery has already been updated in the current time step, preventing self-discharge from being applied twice during dispatch.
The total initial capital cost is:
\begin{align*}
    IC_{total} = & \; IC_{PV} + IC_{WT} + IC_{BS} + IC_{DG} + IC_{conv} \\
    = & \; IC_{PV,\; kW} \cdot P_{PV,\;r,\;total} + IC_{WT,\;kW}\cdot P_{WT,\;r,\;total} + IC_{BS,\;kWh} \cdot E_{b,\; init} + IC_{DG,\;kW}\cdot P_{DG,\;r,\;total} + IC_{conv} \\
    = & \; IC_{PV,\; kW} \cdot n_s \cdot P_{PV,\;r} + IC_{WT,\;kW}\cdot n_w \cdot P_{WT,\;r} + IC_{BS,\;kWh} \cdot E_{b,\; init} + IC_{DG,\;kW}\cdot P_{DG,\;r,\;total} + IC_{conv}
\end{align*}
The total annual recurring costs are:
\begin{align*}
    C_{rec} & = OM_{fix,\;PV} + OM_{fix\;WT} + OM_{fix,\;BS} + OM_{fix,\;DG}   \\
    & + OM_{var, \; DG,\; annual} + C_{fuel,\; DG,\; annual} + SUC_{DG,\;annual} + SDC_{DG,\;annual}  \\
\end{align*}
where
\begin{align*}
    OM_{var, \; DG,\; annual} & = \Sigma_{t=1}^{t=8760} \; DG_{ON}(t) \cdot OM_{var, \; DG}, \; C_{fuel,\; DG,\; annual} & = \Sigma_{t=1}^{t=8760} \; C_{fuel,\; DG}(t). 
\end{align*}
and $SUC_{DG,\;annual}$ and $SDC_{DG,\;annual}$ increase by $SUC_{DG}$ and $SDC_{DG}$ at startup and shutdown, respectively, when $DG_{ON}(t) \neq DG_{ON}(t-1)$. First-year recurring costs are assumed to repeat throughout the project. Their present worth (PW) is calculated as \cite{lcc,ajan2003}:
\begin{equation}
    PW_{C_{rec}} = C_{rec} \frac{\left(\frac{1+f}{1+i}\right)\left(\left(\frac{1+f}{1+i}\right)^{t_{sys}} - 1\right)}{\left(\frac{1+f}{1+i}\right)- 1}
\end{equation}
The total non-recurring costs arise from periodic replacement of the DG and BS:
\begin{equation}
    C_{non-rec} = RC_{DG} + RC _{BS}
\end{equation}
Their present worth is calculated as \cite{lcc,ajan2003}:
\begin{equation}
    PW_{C_{non-rec}} = C_{non-rec} \frac{\left(\frac{1+f}{1+i_{adj}}\right)\left(\left(\frac{1+f}{1+i_{adj}}\right)^{t_{sys}} - 1\right)}{\left(\frac{1+f}{1+i_{adj}}\right)- 1}
\end{equation}
where $i_{adj}$ is an adjusted interest rate:
\begin{equation}
    i_{adj} = \frac{(1+i)^L}{(1+f)^{L-1}} - 1
\end{equation}
Here, $L$ is the replacement interval in years. Assuming that operation repeats the first year, the battery replacement interval is $L_{BS} = \frac{t_{BS}}{n_{cycles}}$. With hourly time steps ($\Delta t = 1 \; h$), summing $DG_{ON}$ gives the DG's annual operating hours, so its replacement interval is $L_{DG} = \frac{t_{DG}}{\Sigma_{t=1}^{t=8760} \; DG_{ON}}$.
    
The total net present cost (TNPC) is the sum of initial costs and the present worth of recurring and replacement costs \cite{kaabeche2011a}: 
\begin{equation}
    TNPC \; [\$] = IC_{total} + PW_{C_{non-rec}} + PW_{C_{rec}}
\end{equation}
The total annualized cost (TAC) is:
\begin{equation}
    TAC = TNPC \cdot CRF 
\end{equation}
where CRF is the capital recovery factor, which converts present worth into an equivalent annual cost \cite{kaviani2009}:
\begin{equation}
    CRF = \frac{r(1+r)^{t_{sys}}}{(1+r)^{t_{sys}} - 1}
\end{equation}
where $r$ is the real interest rate:
\begin{equation}
\label{eq:real_interest}
    r = \frac{i-f}{1 + f}
\end{equation}

\subsection{Dispatch optimization problem costs}

Daily dispatch costs cover only the operating costs incurred over the 24~h horizon, such as fuel and O\&M, and exclude capital and replacement costs, so no discounting is needed:
\begin{align*}
    C_{daily} & = C_{fuel,\; DG,\; daily} + OM_{var, \; DG,\; daily} + SUC_{DG,\;daily} + SDC_{DG,\;daily} \\
    & + \frac{OM_{fix,\;PV} + OM_{fix,\;WT} + OM_{fix,\;BS} + OM_{fix,\;DG}}{365} \\
    OM_{var, \; DG,\; daily} & = \Sigma_{t=1}^{t=24} \; DG_{ON}(t) \cdot OM_{var, \; DG}, \; C_{fuel,\; DG,\; daily} = \Sigma_{t=1}^{t=24} \; C_{fuel,\; DG}(t). 
\end{align*}
In the dispatch optimization of \cref{sec:dispatch_opt}, a battery wear cost of 0.068~USD per kWh discharged is added to $C_{daily}$.

\subsection{Objective functions}

Sizing and dispatch pursue several objectives that can conflict, so we solve a multiobjective, multiperiod, constrained and bounded problem, which also exposes the trade-offs between the objectives. The five objectives are defined below.

\subsection{Levelized cost of electricity (LCOE)}
$LCOE$ [USD/kWh] is the total annualized cost divided by the annual electricity demand:
\begin{align}
    LCOE & = \frac{TAC}{\Sigma_{t=1}^{t=8760} \; E_{load}(t)} = \frac{TAC}{\Sigma_{t=1}^{t=8760} \; P_{load}(t)\cdot \Delta t} = \frac{TAC}{\Sigma_{t=1}^{t=8760} \; P_{load}(t)}
\end{align}
where $\Sigma_{t=1}^{t=8760} \; E_{load}(t)$ is the annual electricity demand [kWh] with hourly steps ($\Delta t = 1$~h). Sizing minimizes LCOE to keep electricity affordable; dispatch minimizes the daily cost of electricity (COE) [USD/kWh] instead:
\begin{align*}
    COE & = \frac{C_{daily}}{\Sigma_{t=1}^{t=24} \; P_{load} (t)\cdot \Delta t}  = \frac{C_{daily}}{\Sigma_{t=1}^{t=24} \; P_{load}(t)}
\end{align*}

\subsection{Emissions \label{sec:norm}}
The emissions objective covers the airborne pollutants emitted by the backup DGs, including particulate matter (PM), volatile organic compounds (VOC) and greenhouse gases (GHG). \cref{tab:emissions} lists the emission factors for diesel in the DE and natural gas in the MT. Total emissions $Em$ [kg] are the sum of these factors multiplied by the energy from fossil sources, $\Sigma_{t=1}^{t=T} \; P_{DG}(t) \cdot \Delta t$, with $T = 24$~h for dispatch and 8,760~h for sizing.

\begin{table}[htbp]
\centering
\begin{tabular}{@{}lcc@{}}
\toprule
\textbf{Parameter} & DE & MT \\ 
\midrule
$CO_2$ [$kg/kWh$] & 0.649 \cite{wu2016} & 0.631 \cite{chp_mt} \\
$CO$ [$g/kWh$] & 4.063 & 2.851\\
$NO_x$ [$g/kWh$] & 18.857 & 20.884\\
$SO_2$ [$g/kWh$] & 0.0074 & 0.003\\
$VOC$ [$g/kWh$] & 1.502 & 0.604\\
$PM$ [$g/kWh$] & 1.338 & 0.047\\
$PM2.5$ [$g/kWh$] & 1.338 & 0.047\\
$PM10$ [$g/kWh$] & 1.338 & 0.047\\
\bottomrule
\end{tabular}
\caption{Emission factors for the DE and MT \cite{emissions}. The DE $SO_2$ factor assumes diesel with 15~ppm sulfur, all converted to $SO_2$ (AP-42 Table 3.4-1~\cite{ap42_34}); at 50~ppm, the limit for diesel in Kenya~\cite{bd_sulphur}, it is 0.025~$g/kWh$, which changes total emissions by less than 0.01\%.}
\label{tab:emissions}
\end{table}

To put LCOE and emissions on a scale comparable to the other three objectives, which lie between 0 and 1, we divide them by the LCOE and emissions of a baseline in which a single DG (a DE or an MT, matching the hybrid system) with rating $P_{DG,\;r} = P_{load,\; peak} \cdot (1+SF)$, $SF \geq 0$, supplies the whole load. The baseline DG runs throughout the study period with one startup and one shutdown, and needs no converter because its AC output supplies the load directly. Its costs, used to compute $LCOE_{base}$, are:

\begin{gather*}
    IC_{base} = IC_{DG}, \; C_{rec,\; base} = OM_{fix,\;DG} + OM_{var, \; DG,\; annual} + C_{fuel,\; DG,\; annual} + SUC_{DG} + SDC_{DG}  \\
    C_{non-rec,\; base} = RC_{DG}, \; C_{daily,\; base} = C_{fuel,\; DG,\; daily} + OM_{var, \; DG,\; daily} + SUC_{DG} + SDC_{DG} + \frac{OM_{fix,\;DG}}{365}
\end{gather*}

\noindent The annual terms $OM_{var, \; DG,\; annual}$ and $C_{fuel,\; DG,\; annual}$ assume 8,760~h of DG operation to meet the annual load. The daily terms $C_{fuel,\; DG,\; daily}$ and $OM_{var, \; DG,\; daily}$ assume 24~h of operation to meet the daily load. $Em_{base}$ likewise assumes that the DG meets all demand. The baseline DG follows the load in every hour, so its annual fuel use is $8760\,\beta_{DE} P_{DG,\;r} + \alpha_{DE} \sum_t P_{load}(t)$; it does not run at rated power for 8,760~h. In every sensitivity case we recompute $LCOE_{base}$ with the same fuel price, interest and inflation rates as the hybrid system, so the normalized LCOE compares the two systems under identical economic conditions, while the baseline DG rating stays at 16~kW so that the normalization does not move with the hybrid system's DG rating. Because the baseline itself moves with fuel price and financial parameters, the normalized LCOE can fall while the absolute LCOE rises; \cref{sec:sizing_sensitivity} therefore reports both.

\subsection{Deficiency of power supply probability (DPSP)}
$DPSP$ is the ratio of unmet load, i.e.\ consumption that is curtailed, interrupted or disrupted, to the total demand over the study period \cite{prasad2006}:
\begin{equation}
    DPSP = \frac{\Sigma_{t=1}^{t=T} \; P_{lost}(t)}{\Sigma_{t=1}^{t=T} \; P_{load}(t)}
\end{equation}
where $T = 8{,}760$~h for sizing and $T = 24$~h for dispatch. DPSP lies between 0 (all load served) and 1 (no load served); minimizing it maximizes reliability.

\subsection{Relative excess or dump power generated (REPG)}

REPG is the ratio of the excess DC power sent to dump loads or grounded to the total DC power generated \cite{kaabeche2011b}:
\begin{align}
    REPG = Dump & = \frac{\Sigma_{t=1}^{t=T} \; P_{dump}(t)}{\Sigma_{t=1}^{t=T} \; P_{gen}(t)} = \frac{\Sigma_{t=1}^{t=T} \; P_{RES}(t) + P_{DG}(t) \cdot \eta_{rec} + P_{BS}(t) - \frac{P_{load}}{\eta_{inv}}}{\Sigma_{t=1}^{t=T} \; P_{RES}(t) + P_{DG}(t) \cdot \eta_{rec}} \label{eq:Pbal}
\end{align}
where the second equality in \cref{eq:Pbal} follows from the power balance. REPG measures how efficiently the system uses its generation; minimizing it ($0 \leq REPG \leq 1$) reduces wasted energy and renewable curtailment.

\subsection{Renewable energy penetration}
The renewable energy fraction (REF) is the share of generation from renewable sources:
\begin{align}
    REF & = \frac{\Sigma_{t=1}^{t=T} \; P_{RES}(t)}{\Sigma_{t=1}^{t=T} \; P_{gen}(t)} = \frac{\Sigma_{t=1}^{t=T} \; P_{PV}(t) + P_{WT}(t) \cdot \eta_{rec}}{\Sigma_{t=1}^{t=T} \; P_{RES}(t) + P_{DG}(t) \cdot \eta_{rec}}
\end{align}
where $T$ is 8,760~h for sizing and 24~h for dispatch. Maximizing REF ($0 \leq REF \leq 1$), or equivalently minimizing $1-REF$, reduces dependence on fossil fuels.

\section{Effects of varying relative objective weights on the sizing results \label{app:weights}}

\begin{figure}[htbp]
  \centering
  \includegraphics[width=\linewidth]{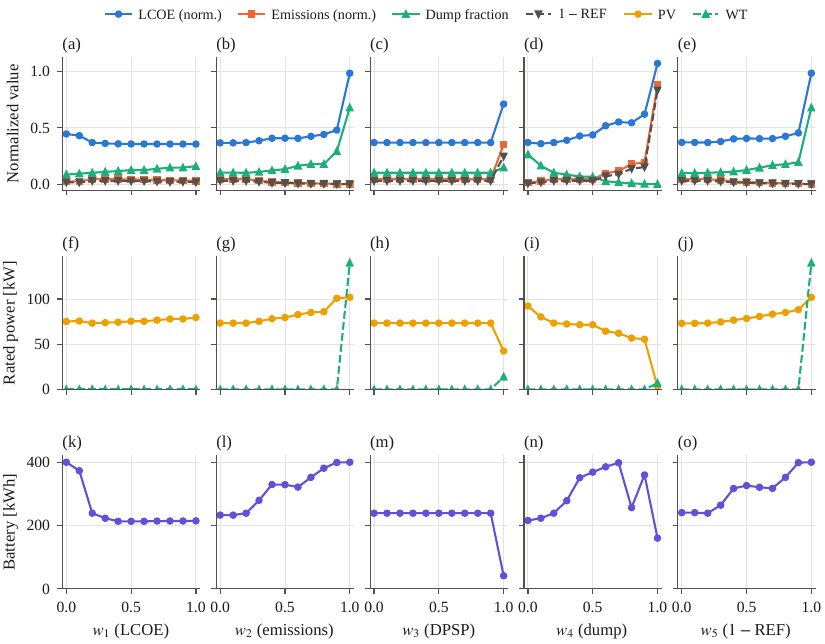}
  \caption{\label{fig:vary_weights} Effect of the objective weights on (a--e) the normalized objectives, (f--j) the rated PV and WT capacity and (k--o) the battery capacity. In each column one weight varies from 0 to 1 and the others share the remainder equally.}
\end{figure}

The weighted-sum design depends on the weights. We varied each weight $w_k$ from 0 to 1 while the other four shared the remainder equally ($w_j = (1-w_k)/4$), re-optimizing the design at each value (best of three integer-coded PSO runs, 16~kW DE). \cref{fig:vary_weights} shows the normalized objectives and the resulting capacities; the equal-weight design ($w_k = 0.2$) is [288, 0, 239] with LCOE 0.184~USD/kWh.

Weighting cost more heavily moves the design little: for $w_1 \geq 0.3$ the optimum stays near [312, 0, 215], with LCOE of 0.177~USD/kWh, close to the equal-weight value. Weighting emissions ($w_2$) or the renewable fraction ($w_5$) enlarges the PV array and battery and cuts DE operation, from 589~h per year at equal weights to 23~h at $w_2 = 0.9$, at an LCOE of 0.239~USD/kWh. At $w_2 = 1$ or $w_5 = 1$ the problem only rewards eliminating diesel, and the optimum moves to the upper bounds ([400, 40, 400]) at 0.490~USD/kWh. Weighting dumped energy ($w_4$) shrinks the renewable array and runs the DE more (2349~h per year at $w_4 = 0.8$); at $w_4 = 1$ the design degenerates to a nearly diesel-only system ([12, 2, 160]). The DPSP weight $w_3$ has no effect below 1 because DPSP is zero for every design with a 16~kW DE; at $w_3 = 1$ the objective is flat and the returned design ([167, 4, 41]) is arbitrary. These extremes show why single-objective weights are a poor basis for design and why \cref{sec:pareto_select} selects designs from the Pareto set instead.

\section{\label{app:improv} Model caveats, potential refinements and extensions}

\subsection{Alternative dispatch strategies}

The sizing rule starts the DG whenever renewable output and the battery cannot cover the load. Running the DG at times of low demand and reserving battery and renewable energy for peak hours would need a different priority order, in which part of the renewable output charges the battery instead of serving the load at once. Other strategies may improve the objectives further, so the space of dispatch strategies deserves exploration at the design stage.

\subsection{Improving computational efficiency}

Nonconvex formulations are harder and slower to solve. Relaxations, variable transformations or linearizations could make the problem easier; for example, replacing the discrete decision variables by continuous ones could convexify the problem and speed up optimization, and could make room for more detailed nonlinear component models that are intractable in a nonconvex setting. Vectorizing the objective function would also shorten run times.

\subsection{More comprehensive multiyear modeling}

The sizing simulation covers one year and assumes that later years repeat it, with discounting. It therefore ignores the 10--15\% annual load growth expected in Kenya and Tanzania from population growth, economic development and rising incomes and living standards \cite{usea}, as well as year-to-year changes in other parameters, such as volatile fuel prices, which the sensitivity analysis covers only one at a time. Dispatch is optimized one day at a time; longer horizons would show how the system responds to multi-day disturbances.

\subsection{Simulation using smaller time steps}

Sub-hourly steps (e.g., $\Delta t = 5$~min) would represent the battery dynamics and SOC more accurately and allow DG ramp-rate limits to be enforced on startup and shutdown. They would also smooth the step changes in the hourly dispatch and SOC profiles. The cost is more computation and the need for sub-hourly climate and load data, which are rarely available for rural African sites.

\subsection{\label{sec:dr} Demand response (DR) and demand-side management (DSM)}

Demand-side management can make the system more robust to load fluctuations. Demand response (DR) programs have been modeled with game theory, mechanism design and auction theory. In one such model \cite{nwulu2017}, the microgrid operator maximizes its net expected benefit:
\begin{equation}
    \max_{x,y} \; V_2 (\theta, \lambda)
\end{equation}
where $V_2$ is the benefit function of the operator:
\begin{equation}
    V_2 (\theta, \lambda) = \lambda x - y
\end{equation}
where $x$ is the load shifted or curtailed by a customer of type $0 \leq \theta \leq 1$, which measures the customer's willingness to shift or curb demand ($\theta = 1$ for the most willing and $\theta = 0$ for the least willing). $\lambda$ is the value of power interruptibility, which measures how much the operator gains from modifying a specific load and is obtained from an optimal power flow (OPF), and $y$ is the compensation paid to the customer for the load shifted or curtailed. The customer's benefit function is:
\begin{equation}
    V_1(\theta,x,y) = y - c(\theta,x)
\end{equation}
where $c(\theta,x)$ is the cost to a customer of type $\theta$ of shifting or reducing load by $x$~kW, increasing in $x$ and decreasing in $\theta$. Constraints cap the operator's compensation budget and each customer's daily interruptible power.

Agent-based formulations of this kind suit small networks such as microgrids, with few market participants, but their computational cost grows quickly with the number of agents and the complexity of the network. They also need two-way communication between all agents, which requires infrastructure upgrades, and smart meters and home energy monitors connected to the Internet of Things (IoT) raise privacy concerns; non-intrusive load monitoring (NILM) and disaggregation can provide enough information for load management while preserving privacy.

The results of such models are also uncertain, because they depend on decisions by consumers, producers and prosumers, who may act irrationally. Each agent's response function, preferences and private costs must be estimated; in the model above, the types $\theta_i$ and cost functions $c_i(\theta_i,x)$ of all participants are hard to determine. The results can often be stated only in probabilistic terms, a further barrier to practical use.

\subsection{Improved load and RES forecasting}

More accurate forecasts of load and of solar and wind output would reduce the effect of uncertainty, especially when combined with predictive control and online optimization that update the forecasts at each time step instead of relying on static profiles or historical data. Sizing and dispatch could also be tested with load profiles from sources other than REopt, such as empirical relations for the demand of isolated rural African villages \cite{load_profile}:
\begin{equation}
    P \; [kW] = e^{[sin(0.3409 - sin(0.68039t) - 0.16801t)]}
\end{equation}


\subsection{Improved control techniques \label{app:improvements}}

The dispatch model assumes perfect foresight: the scheduler knows the actual demand and renewable output over the whole planning horizon from precalculated load profiles and historical climate data. This is useful for sizing, for designing nominal dispatch strategies and as an idealized benchmark for other strategies \cite{parisio2014}, but such reliable forecasts are rarely available in practice. Model predictive or receding-horizon control (\cref{app:control}) can improve operation under uncertainty. Our objective is a nonconvex, non-smooth black-box function without a closed form or gradient, which makes it harder to cast in standard MPC form; commercial solvers such as GUROBI, however, solve mixed-integer linear and quadratic MPC problems efficiently without complicated heuristics or decomposition \cite{parisio2014}.

\section{Detailed survey of optimization algorithms \label{app:opt_alg}}

Besides using tools such as HOMER Pro and DER-CAM, many studies build their own sizing and dispatch models, usually as nonlinear mixed-integer programs. Relaxations and variable transformations can turn these into mixed-integer linear programs, which are more tractable than quadratic programs, and assumptions can be adjusted to convexify the problem; for example, quadratic DG fuel-cost functions are often approximated by the maximum of affine functions \cite{parisio2014}. The subsections below describe common global optimization methods for such problems. These artificial-intelligence search techniques tend to outperform classical methods such as linear programming as dispatch problems grow more complex \cite{Qu2018}; Lyapunov optimization \cite{pathak2017} and discrete Fourier transforms \cite{dft} have also been used.

\subsection{Genetic (GA) and evolutionary algorithms (EA)}

Genetic algorithms solve smooth or non-smooth problems with any type of constraint, including integrality. They are stochastic, population-based methods that search the space of candidate solutions by \textit{selection}, \textit{mutation} and \textit{crossover} among population members, imitating evolution by natural selection, with each member representing a candidate solution \cite{nemati2015}. The population of chromosomes is initialized as vectors of bits or character strings. The algorithm minimizes a fitness value that measures deviation from the objectives and constraints, for example a weighted sum of the cost, penalties on constraint violations and a parity degree that measures the similarity among individuals (\cref{eq:GA_app}), where $W_{Pen}$ and $W_{Par}$ are weighting factors and $R_{Cost/Pen}$ and $R_{Cost/Par}$ are ratios of mean values that make the three terms comparable. The parity term rewards diversity: it raises the survival probability of individuals that differ from the group and so reduces the risk of converging to, and stalling at, a local optimum \cite{nemati2015}.

\begin{equation}
    F = Cost + W_{Pen} \cdot R_{Cost/Pen} \cdot Penalty + W_{Par} \cdot R_{Cost/Par} \cdot Parity
    \label{eq:GA_app}
\end{equation}

At each iteration the fitness of every chromosome is evaluated and the best solutions are selected \cite{pathak2017}; these are modified and recombined over many generations until a stopping criterion such as a generation limit is met \cite{nemati2015}. Evolutionary programming rests on the same Darwinian principles, but where GA simulates micro-level processes such as chromosomes, evolutionary algorithms mimic macro-level processes such as phenotype variation and inheritance \cite{pathak2017}. Because they evolve a population, these methods can find multiple well-distributed non-dominated solutions in a single run with little information about the search space \cite{Qu2018}. \textit{Differential evolution} is a variant that differs in the selection step, using non-domination sorting and ranking \cite{Qu2018}; it has solved environmental/economic dispatch (EED) and unit commitment (UC) problems effectively \cite{lu2011}.

Multiobjective problems can be solved by combining the objectives into a weighted sum or by approximating the Pareto set, the set of feasible non-dominated solutions that cannot be improved in one objective without worsening another. Genetic and evolutionary methods are well suited to generating such sets and to showing the trade-offs between objectives such as cost, efficiency and reliability \cite{fadaee2012}.

Their main disadvantage is that complexity, and with it the difficulty of programming and solving, grows quickly with the number of parameters \cite{fathima2015}. Standard GA explores new regions of the search space only to a limited extent, and its crossover and mutation operators, preset from experience, adapt poorly; improved adaptive genetic algorithms avoid premature local convergence and reach global optima faster and more reliably \cite{wu2016}.

\subsection{Particle swarm optimization (PSO) \label{app:pso}}

Particle swarm optimization (PSO) mimics the collective behavior of a swarm searching for food within a region \cite{fathima2015}, as seen in animal societies and insect colonies \cite{pathak2017}. It is an iterative, agent-based method in which a population of particles, each a candidate solution, explores the space of feasible solutions \cite{alrashidi2010}. PSO is easy to implement, has few parameters but it can stall in local optima and has difficulty satisfying equality constraints when the swarm is initialized and updated \cite{wu2014}. Improved multiobjective PSO variants have sized microgrid batteries and solved a range of EED problems, including large non-convex problems such as optimal power flow (OPF) \cite{alrashidi2010}, converging quickly to good solutions at low computational cost \cite{pathak2017}.

Standard PSO initializes the particles at random positions $[x_{i1}, x_{i2}, \dots, x_{id}]$ with velocities $[v_{i1}, v_{i2}, \dots, v_{id}]$, where $i = 1, 2, \dots, n$ indexes the $n$ particles and $d$ is the dimension of the search space. Each particle is drawn toward two points: the best position it has found so far ($pbest$) and the best position found by the whole swarm ($gbest$) \cite{borhanazad2014}. At each iteration:

\begin{enumerate}
    \item Evaluate the fitness of each particle.
    \item Update both $pbest$ and $gbest$.
    \item Update the velocity and position of each particle using \cite{wu2014}:
    \begin{align}
    \label{eq:pso}
        v_{i,d}^{k+1} & = \omega \cdot v_{i,d}^{k} + c_1 \cdot rand_1^k \cdot (pbest_{i,d}^k - x_{i,d}^{k}) + c_2 \cdot rand_2^k \cdot (gbest_{i,d}^k - x_{i,d}^k) \\
        x_{i,d}^{k+1} & = x_{i,d}^{k} + v_{i,d}^{k+1}
    \end{align}
\end{enumerate}

The velocity update in \cref{eq:pso} is a randomly weighted sum of three terms (inertia, attraction to $pbest$ and attraction to $gbest$), where $v_{i,d}^k$ and $x_{i,d}^k$ are the velocity and position of particle $i$ along dimension $d$ in iteration $k$, $\omega$ is the inertia weight, $c_1$ and $c_2$ are learning factors, $pbest_{i,d}^k$ and $gbest_{i,d}^k$ are the local and global best values of coordinate $d$ in iteration $k$, and $rand_1^k$ and $rand_2^k$ are random numbers drawn uniformly from $(0,1)$ \cite{wu2014}. The updates continue until a termination condition, such as an iteration limit or a target fitness, is met, and $gbest$ is returned as the solution. Related swarm-intelligence methods include ant and bee colony optimization, bacterial foraging and group search optimizers \cite{pathak2017}, all part of the wider family of bio-inspired algorithms for complex multiobjective problems.

\subsection{Surrogate optimization}

Surrogate optimization targets objective functions that are expensive to evaluate. It samples quasirandom points within the bounds, interpolates the true objective at these points with a radial basis function to build a cheaper surrogate \cite{gutmann2001}, and takes the minimizer of the surrogate as an approximation of the true minimizer. It balances two goals: exploring the search space for a global minimum and reaching a good solution in few evaluations. It works best on smooth, continuous objectives, although smoothness is not required, and it converges to the global optimum on bounded domains, though slowly \cite{gutmann2001}. It can greatly speed up sizing and dispatch when each evaluation is expensive; when evaluations are cheap, as in our 8,760-hour simulation, fitting the surrogate dominates the run time (\cref{tab:bench_stats}). No stopping criterion reliably detects that the solver is near a global optimum, so it stops after reaching a limit on iterations, evaluations or elapsed time and returns the best point found within that budget.

\subsection{Simulated annealing (SA)}

Simulated annealing (SA) is a stochastic search modeled on the annealing of metals, which are melted and then cooled slowly into low-energy crystalline structures; the temperature schedule is the key control. At a fixed temperature $T$, the current state is perturbed to a new state with energy difference $\Delta E$. The new state is accepted if $\Delta E < 0$; otherwise it is accepted with probability:
\begin{equation}
\label{eq:sa}
    Pr\;(acceptance) = exp\left(\frac{-\Delta E}{k_BT}\right)
\end{equation}
where $k_B$ is the Boltzmann constant. Perturbations repeat until the system reaches equilibrium at temperature $T$, after which the temperature is lowered slightly and the sequence repeats. In microgrid optimization the energy $E$ becomes the objective, the temperature $T$ a control parameter, and \cref{eq:sa} the probability of accepting a candidate system configuration or dispatch strategy \cite{fathima2015}. SA is less common than GA, EA and PSO for the sizing and dispatch of hybrid systems; one study found it faster to converge but less efficient than tabu search for sizing that minimizes LCOE \cite{sa_tabu}.


\section{Detailed survey of microgrid control strategies \label{app:control}}

Beyond design and dispatch scheduling, a third area of research concerns the real-time control of microgrids. Closed-loop control is likely to outperform open-loop, model-based scheduling in practice, because it uses measurements of the actual system state to correct for variations in demand and supply \cite{tazvinga2014}. Control is needed to manage variability in renewable output, and more so as localized, distributed generation grows, especially in islanded and off-grid systems, where power quality, safety and overloads are particular concerns \cite{pourbabak2017}; bidirectional flows of power and information add complexity. Feedback control therefore improves robustness to forecast errors in load and renewable output and to disturbances such as wind lulls and other extreme weather, natural disasters and sudden demand spikes from heating or cooling loads. Microgrid control is conventionally organized in three levels. Primary control regulates the voltage, frequency and power of individual converters. Secondary control improves power quality, mitigating voltage or frequency deviations, imbalances and harmonics; in practice it restores frequency and voltage to their nominal values after primary control by adjusting real and reactive power setpoints in real time. Tertiary control handles power exchanges with the main grid or with adjacent microgrids \cite{dragicevic_2017}, and can include system-wide techno-economic optimization at the design stage and efficiency improvements in day-to-day dispatch \cite{mg_review}. Control architectures for large cyber-physical systems such as microgrids are centralized, decentralized or distributed.

\subsection{Centralized control}

Centralized control is the most mature of the three architectures and has long been used to control and economically dispatch generators in power systems \cite{su2015}. A single controller collects and processes data from the whole system and sends control signals to each agent through supervisory control and data acquisition (SCADA), an automation system that centrally manages the control, data gathering and monitoring of a power system \cite{scada}. The framework is conceptually simple and has been studied for decades, producing many controllers, particularly for linear systems \cite{farina2012}. It is hard to implement in practice, however, because it needs extensive two-way communication that grows with the number of agents; real-time transmission of measurements and computation of control actions may be infeasible over a large area, so cost and computation scale poorly with network size, and offline model-based synthesis is difficult for systems with many inputs, states and outputs. It also requires all agents to share private information with the central controller, which may be unacceptable for socioeconomic and political reasons \cite{farina2012}. Centralized control is nevertheless a good starting point for small microgrids with few producers, consumers and prosumers and simple topologies, and can be replaced by a distributed scheme as the system grows.

\subsection{Decentralized control}

In decentralized control, each agent, node or subsystem (e.g., a DG, consumer or battery) has its own controller and acts only on local measurements such as voltage or frequency \cite{pourbabak2017}. No global information is needed. Agents retain their information locally, sharing it neither with a central unit nor with other nodes \cite{gan2012}. Control variables are computed in parallel, and real-time communication is easier because sensors, controllers and actuators within a subsystem are close together \cite{farina2012}. Without communication between controllers, however, performance is limited and global optimality or stability cannot be guaranteed \cite{pourbabak2017}. Both centralized and decentralized schemes have improved transient stability in small power systems such as microgrids \cite{senjyu2005}, and decentralized systems can be more robust than comparable centralized ones: if some nodes disconnect because of faults, the others keep operating on local information.

\subsection{Distributed control}

Distributed control is consensus-based: each agent uses local measurements and information shared with its nearest neighbors through two-way communication. This enables global optimization with efficient parallel computation, lets the designer trade performance against communication burden and adapt the network design \cite{farina2012}, and leaves agents some control over their privacy. Distributed control thus combines strengths of the centralized and decentralized schemes and is increasingly popular for microgrids. Because information is shared locally, it copes with the changing structure of the microgrid and of future smart grids, and it meets the \textit{peer-to-peer} and \textit{plug-and-play} requirements. Peer-to-peer operation means that no single component is critical, so the system keeps working when an individual generator or storage unit fails; plug-and-play means that a new unit can be installed anywhere without reconfiguring the network or controller \cite{mg_prop}. The microgrid can then grow as producers and customers connect \cite{pourbabak2017}, which suits remote rural electrification. Its scalability, fast response and model independence also suit real-time DR and DSM algorithms that engage users and generators \cite{deng2014} and systems with flexible DGs and responsive loads \cite{dist_1, dist_2}. Open challenges are the design of the communication and control topology, infrastructure upgrades for bidirectional flows, and fast convergence to consensus among local agents \cite{pourbabak2017}.

\subsection{Model predictive control \label{app:mpc}}

Model predictive control (MPC) is one of the most widely used methods for optimal microgrid operation. It turns the static optimization into a dynamic one by solving, at each sampling instant, an open-loop optimal control problem over a receding finite horizon from the current state; only the first control action is applied, and the process repeats from the new state \cite{mpc}. Linear MPC solves a quadratic (QP) or linear program (LP) at each control interval, i.e.\ a quadratic or linear cost subject to linear constraints. Unlike PI and PID controllers, or control laws synthesized offline as a precomputed feedback policy for all states, MPC solves the optimal control problem online for the \textit{current} state \cite{mpc}. It can therefore anticipate future events, such as fluctuations in renewable supply or changes in demand, and act to damp their effects while satisfying time-varying operating constraints \cite{parisio2014}, which improves performance over periods in which disturbances occur.

Predictive control has been implemented with mixed-integer linear and nonlinear programs, because microgrid models need both discrete variables (e.g., the on/off states of DGs) and continuous ones (e.g., battery SOC and charge and discharge rates) \cite{parisio2013, parisio2014}. Such feedback compensates for uncertain renewable output and time-varying loads. Robust linear and nonlinear economic MPC (E-MPC) methods optimize dispatch setpoints to minimize operating cost \cite{zachar2016} or to maximize profit from selling power to the main grid \cite{pereira2017}; finite-horizon MPC objectives comprise stage, terminal and penalty costs, and E-MPC uses economic costs as the stage cost. Closed-loop discrete linear MPC has been applied to a PV--wind--diesel--battery system similar to the one studied here \cite{tazvinga2014}. Predictive control has also handled uncertainty in generator model parameters \cite{prodan2014}, and adaptive, switched MPC uses switching constraints, rather than switched state-space models, to represent transitions such as between battery charging and discharging \cite{zhu2015}.


\section{Detailed survey of microgrid design tools \label{app:tools}}

\subsection{HOMER Pro}

HOMER (Hybrid Optimization of Multiple Energy Resources), also known as Hybrid Optimization Model for Electric Renewables Pro, was developed by the National Renewable Energy Laboratory (NREL) and is distributed by HOMER Energy. It is widely used in the literature surveyed here for microgrid design, from village power systems and island utilities to grid-connected campuses and military bases. HOMER evaluates combinations of technologies through simulation and uses derivative-free optimization and exhaustive search \cite{weng2016} to minimize total net present cost (TNPC), the present value of costs minus revenues. It takes six types of input: climate data, load profiles, equipment characteristics, search-space specifications, economic data and technical data \cite{bahramara2016}. Component sizes are determined through simulation, optimization and sensitivity analysis, with time steps from 1 minute to 1 hour to capture intraday variability. Sensitivity analysis assesses how fuel prices, wind speed, solar radiation and component costs affect the lowest-cost feasible design.

\subsection{DER-CAM}

The Distributed Energy Resources Customer Adoption Model (DER-CAM), developed by Lawrence Berkeley National Laboratory (LBNL), optimizes both design and dispatch for DER investments in buildings and hybrid multi-energy microgrids. It allows objective functions to be modified and supports multiobjective optimization. DER-CAM uses a mixed-integer linear program (MILP) \cite{dercam} with inputs similar to those of HOMER. It represents loads and solar irradiance through characteristic day types that capture seasonal variation \cite{weng2016}. Outputs include the selected technologies, capacities, locations and dispatch schedules needed to meet economic, resilience and reliability targets, with cost and emissions breakdowns by end use. The default model assumes no deterioration in component output or efficiency over time and excludes startup, shutdown and ramp-rate constraints \cite{dercam}.

\subsection{REopt}

REopt is an MILP-based techno-economic decision-support model from NREL for decentralized energy systems, including grid-connected and off-grid microgrids. It uses hourly or sub-hourly time steps to select the technology mix, capacities and operating schedules for renewables, thermal generation and storage. Objectives include cost savings in grid-connected operation and longer autonomy during islanded operation \cite{tools_summary}. REopt is free to use through a web tool for single-site analysis and an API for multisite analysis, and its code is open source \cite{reopt_news}. We use supporting NREL datasets from REopt cost models to estimate load profiles and examine LCOE for rural Africa \cite{nrel_data}. In NREL's sample comparison, a PV--battery system costs more than a diesel-only system, while adding diesel backup to the PV--battery system gives the lowest LCOE of the three. This supports evaluating hybrid designs that combine renewables, storage and fossil-fueled backup for isolated microgrids.

\subsection{MDT}

The Microgrid Design Toolkit (MDT), developed by Sandia National Laboratories, uses MILP and genetic algorithms. Its discrete-event simulation represents events such as load changes, fuel outages and equipment failures \cite{tools_summary}. MDT supports early design decisions by evaluating alternative microgrid configurations and their performance against user-defined criteria, returning a Pareto front of design options.

\bibliographystyle{unsrt} 

\end{document}